\documentclass[showpacs,preprintnumbers,amsmath,amssymb,superscriptaddress]{revtex4}

\usepackage{calc}
\usepackage{graphicx}
\usepackage{bm}

\begin{document}

\title{Mean first-passage time of surface-mediated diffusion in spherical domains}

\author{O. B\'enichou}
\affiliation{Laboratoire de Physique Th\'eorique de la Mati\`ere Condens\'ee
(UMR 7600), case courrier 121, Universit\'e Paris 6, 4 Place Jussieu, 75255
Paris Cedex}

\author{D. S. Grebenkov}
\affiliation{Laboratoire de Physique de la Mati\`ere Condens\'ee
(UMR7643), CNRS -- Ecole Polytechnique, F-91128 Palaiseau Cedex France} 

\affiliation{Laboratoire Poncelet (UMI 2615), CNRS -- Independent University of Moscow,
Bolshoy Vlasyevskiy Pereulok 11, 119002 Moscow, Russia}

\author{P. E. Levitz}
\affiliation{Laboratoire de Physique de la Mati\`ere Condens\'ee
(UMR7643), CNRS -- Ecole Polytechnique, F-91128 Palaiseau Cedex France}

\author{C. Loverdo}
\affiliation{Laboratoire de Physique Th\'eorique de la Mati\`ere Condens\'ee
(UMR 7600), case courrier 121, Universit\'e Paris 6, 4 Place Jussieu, 75255
Paris Cedex}

\author{R. Voituriez}
\affiliation{Laboratoire de Physique Th\'eorique de la Mati\`ere Condens\'ee
(UMR 7600), case courrier 121, Universit\'e Paris 6, 4 Place Jussieu, 75255
Paris Cedex}

\date{\today}

\begin{abstract}
We present an exact calculation of the mean first-passage time to a
target on the surface of a 2D or 3D spherical domain, for a molecule
alternating phases of surface diffusion on the domain boundary and
phases of bulk diffusion.  The presented approach is based on an
integral equation which can be solved analytically.  Numerically
validated approximation schemes, which provide more tractable
expressions of the mean first-passage time are also proposed.  In the
framework of this minimal model of surface-mediated reactions, we show
analytically that the mean reaction time can be minimized as a
function of the desorption rate from the surface.
\end{abstract}


\maketitle

\section{Introduction}

The kinetics of many chemical reactions is influenced by the transport
properties of the reactants that they involve \cite{Rice:1985a,Hanggi:1990a}. In fact, schematically, any
chemical reaction requires first that a given reactant A meets a
second reactant B.  This first reaction step  can be rephrased
as a search process involving a searcher A looking for a target B.  In
a very dilute regime, exemplified by biochemical reactions in
cells \cite{Alberts:2002} which sometimes involve only a few copies of
reactants, the targets B are sparse and therefore hard to find in this
search process language.  In such reactions, the first step of search
for reactants B is therefore a limiting factor of the global reaction
kinetics.  In the general aim of enhancing the reactivity of
chemical systems, it is therefore needed to optimize the efficiency of
this first step of search.

Recently, it has been shown that intermittent processes, combining
slow diffusion phases with a faster transport, can significantly
increase reactions rates \cite{package_animaux,Obenichou:2008}.  A minimal model demonstrating the
efficiency of this type of search, introduced to account for the fast
search of target sequences on DNA by proteins \cite{Coppey:2004} is as follows (see also \cite{Berg:1981,Slutsky:2004a,Lomholt:2005,Eliazar:2007a}).  The
pathway followed by the protein, considered as a point-like particle,
is a succession of 1D diffusions along the DNA strand (called sliding
phases) with diffusion coefficient $D_1$ and 3D excursions in the
surrounding solution.  The time spent by the protein on DNA during
each sliding phase is assumed to follow an exponential law with
dissociation rate $\lambda$.  In this minimal model, the 3D excursions
are uncorrelated in space, which means that after dissociation from
DNA, the protein will rebind the DNA at a random position
independently of its starting position.  Assuming further that the
mean duration of such 3D excursions $\tau_2$ is finite, it has been
shown that the mean first-passage time at the target can be minimized
as a function of $\tau_1=\lambda^{-1}$, as soon as the mean time spent
in bulk excursions is not too long.  Quantitatively, this condition
writes in orders of magnitude as $\tau_2\leq L^2/D_1$, and the minimum
of the search time is obtained for $\tau_1 \simeq \tau_2$ in the large
$L$ limit.  Note that in this minimal model, where the time $\tau_2$
is supposed to be a fixed exterior parameter, bulk phases are always
beneficial in the large $L$ limit (i.e. allow one to decrease the search
time with respect to the situation corresponding to 1D diffusion
only).

In many practical situations however, the duration of the fast bulk
excursions strongly depends on the geometrical properties of the
system \cite{Benichou:2005a,package_Levitz,Chechkin:2009,Loverdo:2009a} and cannot be treated as an independent variable as assumed in
the mean-field (MF) model introduced above.  An important generic
situation concerns the case of confined systems \cite{Astumian:1985,Adam:1968,Sano:1981}, involving transport
of reactive molecules both in the bulk of a confining domain and on
its boundary, referred to as surface-mediated diffusion in what
follows.  This type of problems is met in situations as varied as
heterogeneous catalysis \cite{bond,Blumen:1984b}, or reactions in porous media and
in vesicular systems \cite{Adam:1968,Sano:1981,Schuss:2007}.  In all
these examples, the duration of bulk excursions is controlled by the
return statistics of the molecule to the confining surface, which
crucially depends on the volume of the system.  This naturally induces
strong correlations between the starting and ending points of bulk
excursions, and makes the above MF assumption of uncorrelated
excursions largely inapplicable in these examples.

At the theoretical level, the question of determining mean
first-passage times in confinement has attracted a lot of attention in recent years  for discrete random walks \cite{Kozak:2002,package_nature,Agliari:2008,Haynes:2008,Reuveni:2010} and continuous processes \cite{Redner:2001a,Sylvain_package,Grebenkov:2007,Condamin:2007yg}.  More precisely, the  surface-mediated diffusion problem considered here generalizes the
so-called narrow escape problem, which refers to the time needed for a
simple Brownian motion in absence of surface diffusion to escape
through a small window of an otherwise reflecting domain. This problem
has been investigated  both in the
mathematical
\cite{Singer:2006b,Schuss:2007,Pillay:2010,Cheviakov:2010} and
physical \cite{Grigoriev:2002a,Benichou:2008,Oshanin:2010,al:2011} literature, partly due to the
challenge of taking into account mixed boundary conditions.  The case
of surface-mediated diffusion brings the additional question of
minimizing the search time with respect to the time spent in
adsorption, in the same spirit as done for intermittent processes
introduced above. The answer to this question is {\it a priori} not
clear, since the mean time spent in bulk excursions diverges for large
confining domains, so that the condition of minimization mentioned
previously cannot be taken as granted, even in the large system limit.
In this context, first results have been obtained in \cite{Benichou:2010}
where, surprisingly enough, it has been found that, even for bulk and
surface diffusion coefficients of the same order of magnitude, the
reaction time can be minimized, whereas MF treatments (see for
instance \cite{Oshanin:2010}) predict a monotonic behavior.

Here, we extend the perturbative results of \cite{Benichou:2010} obtained
in the small target size limit.  Relying on an integral equation
approach, we provide an exact solution for the mean FPT, both for 2D
an 3D spherical domains, and for any spherical target size.  We also
develop approximation schemes, numerically validated, that provide
more tractable expressions of the mean FPT.

\section{The model}

\begin{figure}[hbt]
\begin{centering}
\includegraphics[width=0.45\textwidth]{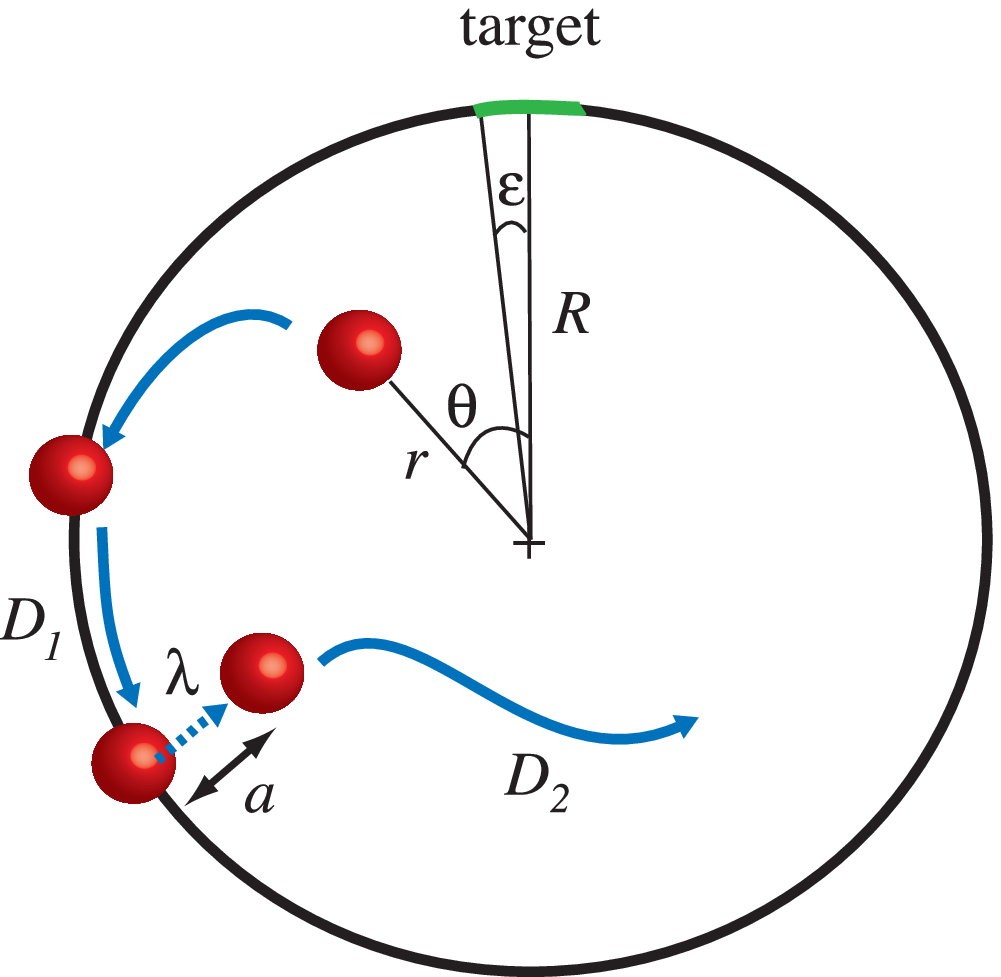}
\caption{\label{fig1}Model}
\end{centering}
\end{figure}

The surface-mediated process under study is defined as follows. We
consider a molecule diffusing in a spherical confining domain of
radius $R$ (see figure \ref{fig1}), alternating phases of boundary
diffusion (with diffusion coefficient $D_1$) and phases of bulk
diffusion (with diffusion coefficient $D_2$).  The time spent during
each one-dimensional phase is assumed to follow an exponential law
with dissociation rate $\lambda$.  At each desorption event, the
molecule is assumed to be ejected at a distance $a$ from the frontier
(otherwise it is instantaneously readsorbed).  Although formulated for
any value of this parameter $a$ smaller than $R$, in most physical
situations of real interest $a\ll R$.  The target is perfectly
absorbing and defined in $2D$ by the arc
$\theta\in[-\epsilon,\epsilon]$, and in $3D$ by the region of the
sphere such that $\theta\in[0,\epsilon]$ where $\theta$ is in this
case the elevation angle in standard spherical coordinates.  Note that
as soon as $\epsilon\neq 0$, the target can be reached either by
surface or bulk diffusion.  In what follows we calculate the mean
first-passage time at the target for an arbitrary initial condition of
the molecule.

\section{2D case}

In this section, the confining domain is a disk of radius $R$ and the
target is defined by the arc $\theta\in[-\epsilon,\epsilon]$.

\subsection{Basic equations}

For the process defined above, the mean first-passage time (MFPT) at
the target satisfies the following backward equations
\begin{eqnarray}
\label{eq:t1}
\frac{D_1}{R^2}t_1''(\theta)+\lambda [t_2(R-a,\theta)-t_1(\theta)] &=& -1 \;\;{\rm for}\;\theta\in[\epsilon,2\pi-\epsilon], \\
\label{eq:t2}
D_2\left ( \frac{\partial ^2}{\partial  r^2} +\frac{1}{r} \frac{\partial }{\partial  r} +
\frac{1}{r^2} \frac{\partial ^2}{\partial  \theta^2}\right) t_2(r,\theta) &=& -1, 
\end{eqnarray}
where $t_1$ stands for the the MFPT starting from the circle at a
position defined on the circumference by the angle $\theta$, and $t_2$
for the MFPT starting from the point $(r,\theta)$ within the disk.  In
these two equations, the first term of the lhs accounts for the
diffusion respectively on the circumference and in the bulk, while the
second term of Eq. (\ref{eq:t1}) describes desorption events.  They
have to be completed by two boundary conditions
\begin{eqnarray}
\label{eq:adsorption}
t_2(R,\theta) &=& t_1(\theta), \\
\label{eq:target} 
t_1(\theta) &=& 0 \;\;{\rm for}  \;\theta\in[0,\epsilon]\cup[2\pi-\epsilon,2\pi], 
\end{eqnarray}
which describe the adsorption events and the absorbing target
respectively.  Eq.(\ref{eq:t2}) is easily shown to be satisfied by the
following Fourier series
\begin{equation}
\label{eq:t2integre}
t_2(r,\theta)=\alpha_0-\frac{r^2}{4D_2}+\sum_{n=1}^\infty \alpha_n r^n \cos(n\theta), 
\end{equation}
with unknown coefficients $\alpha_n$ to be determined.  In particular,
we aim at determining the search time $\langle t_1 \rangle$, defined
as the MFPT, with an initial position uniformly distributed on the
boundary of the confining domain.  Taking Eq.(\ref{eq:t2integre}) at
$r=R$, we have
\begin{equation}
\alpha_0-\frac{R^2}{4D_2}+\sum_{n=1}^\infty \alpha_n R^n \cos(n\theta) = \begin{cases} t_1(\theta)  & 
\text{if $\theta\in[\epsilon,2\pi-\epsilon]$,} \\
0 &\text{if $\theta\in[0,\epsilon]\cup[2\pi-\epsilon,2\pi]$,}
\end{cases}
\end{equation}
so that 
\begin{equation}
\label{eq:defFourier}
\begin{split}
\alpha_0-\frac{R^2}{4D_2} &= \frac{1}{2\pi} \int_\epsilon^{2\pi-\epsilon} t_1(\theta){\rm d}\theta \equiv\langle t_1 \rangle,\\
R^n\alpha_n & =\frac{1}{\pi} \int_\epsilon^{2\pi-\epsilon} t_1(\theta) \cos(n\theta) {\rm d}\theta \hskip 5mm (n\ge1). \\
\end{split}
\end{equation}
In what follows we will make use of the following quantities:
\begin{equation}
\omega\equiv R\sqrt{\lambda/D_1},
\end{equation}
\begin{equation}
x\equiv1-\frac{a}{R},
\end{equation}
and
\begin{equation}
\label{defT}
T\equiv\frac{1}{\lambda}+\frac{R^2 - (R-a)^2}{4D_2}.
\end{equation}

As we proceed to show, two different approaches can be used to solve
this problem.  (i) The first approach, whose main results have been
published in \cite{Benichou:2010}, uses the explicit form of the Green
function for the two-dimensional problem and relies on a small target
size $\epsilon$ expansion.  We recall these perturbative results below
for the sake of self-consistency and give details of the derivation in
Appendix \ref{sec:approach}.  (ii) The second approach presented next
relies on an integral equation which can be derived for $t_1$, and
leads to an exact non-perturbative solution.

\subsection{Perturbative approach} 

It is shown in Appendix that the Fourier coefficients of
$t_2(r,\theta)$ as defined in Eq.(\ref{eq:t2integre}) satisfy an
infinite hierarchy of linear equations, which lead to the following
small $\epsilon$ expansion:
\begin{equation}
\label{eq:alphanperturbation0}
\begin{split}
\alpha_0 & = \frac{R^2}{4D_2} + \omega^2T \left\{ \left(2\sum_{m=1}^\infty\frac{1}{\omega^2\left(1-x^m\right)+m^2}\right)
-\pi \epsilon+ \left(1+2\omega^2\sum_{m=1}^\infty \frac{1-x^m}{\omega^2\left(1-x^m\right)+m^2}\right)\epsilon^2\right\}+\dots , \\
\alpha_n & = \frac{\omega^2 T}{R^n(\omega^2 (1-x^n) + n^2)}\left\{-2+n^2\epsilon^2+\dots  \right\} . \\
\end{split}
\end{equation}
Note that Eq.(\ref{eq:alphanperturbation0}) gives in particular the
first terms of the perturbative expansion of the search time $\langle
t_1 \rangle$ defined in (\ref{eq:defFourier}) and given in
\cite{Benichou:2010}.  It should be stressed that since the coefficients
of $\epsilon^k$ of this expansion diverge with $\omega$, in practice
one finds that the range of applicability in $\epsilon$ of this
expansion is wider for $\omega$ small.

\subsection{Integral equation for $t_1$} 

In this section, we first show that the resolution of the coupled PDEs
(\ref{eq:t1}, \ref{eq:t2}) amounts to solving an integral equation for
$t_1$ only.  As we proceed to show, this integral equation can be
solved exactly.  Writing Eq. (\ref{eq:t1}) as
\begin{equation}
\frac{\partial^2t_1}{\partial \theta^2}  = -\frac{R^2}{D_1}-\omega^2[t_2(R-a,\theta)-t_2(R,\theta)],
\end{equation}
and expanding its right-hand side into a Taylor series leads to
\begin{equation}
\frac{\partial^2t_1}{\partial \theta^2} = -\frac{R^2}{D_1}-
\omega^2 \sum_{k=1}^\infty \frac{(-a)^k}{k!}\left(\frac{\partial ^k t_2}{\partial r^k}\right)_{R,\theta}.
\end{equation}
Substituting the Fourier representation (\ref{eq:t2integre}) for $t_2$
into this equation yields
\begin{eqnarray}
\frac{\partial^2t_1}{\partial \theta^2} = -\frac{R^2}{D_1}-\omega^2\left(\frac{aR}{2D_2}-\frac{a^2}{4D_2}\right)
-\omega^2 \sum_{k=1}^\infty \frac{(-a)^k}{k!}\sum_{n=k}^\infty \alpha_n n(n-1)\dots (n-k+1) R^{n-k}\cos(n \theta).
\end{eqnarray}
Changing the order of summations over $n$ and $k$, using the binomial
formula and the expression (\ref{eq:defFourier}) for $\alpha_n$ give
\begin{eqnarray}
\label{eq:integrodiff_2D}
\frac{\partial^2t_1}{\partial \theta^2}  = -\frac{R^2}{D_1}-\omega^2\left(\frac{aR}{2D_2}-\frac{a^2}{4D_2}\right)
-\frac{\omega^2}{\pi}\sum_{n=1}^\infty (x^n-1) \cos(n \theta)  \int_\epsilon^{2\pi-\epsilon} \cos(n\theta') t_1(\theta') {\rm d}\theta'.
\end{eqnarray}
This integro-differential equation for $t_1$ can actually easily be
transformed into an integral equation for $t_1$, by integrating
successively two times, which leads to 
\begin{equation}
\begin{split}
t_1(\theta) & = \frac{1}{2}\left(\frac{R^2}{D_1}+\omega^2\left(\frac{aR}{2D_2}-\frac{a^2}{4D_2}\right)\right)
(\theta-\epsilon)(2\pi-\epsilon-\theta) \\
& + \frac{\omega^2}{\pi} \sum_{n=1}^\infty (x^n-1) \frac{\cos(n\theta)-\cos(n\epsilon)}{n^2}
\int_\epsilon^{2\pi-\epsilon} \cos(n\theta') t_1(\theta') {\rm d}\theta', \\
\end{split}
\end{equation}
or equivalently to
\begin{equation}
\label{eq:inteq_2d}
\psi(\theta)=(\theta-\epsilon)(2\pi-\epsilon-\theta)+\Omega\sum_{n=1}^\infty(x^n-1)
\frac{\cos(n\theta)-\cos(n\epsilon)}{n^2}\int_\epsilon^{2\pi-\epsilon}\cos(n\theta')\psi(\theta'){\rm d}\theta',
\end{equation}
where
\begin{equation}
\label{eq:psi_2D}
\psi(\theta)\equiv \frac{2 t_1(\theta)}{\omega^2 T} ,
\end{equation}
with $T$ defined in Eq. (\ref{defT}) and $\Omega\equiv
\frac{\omega^2}{\pi}$.  Note that Eq. (\ref{eq:inteq_2d}) holds for
$\theta\in[\epsilon, 2\pi-\epsilon]$. When there is no desorption
(i.e., $\lambda = 0$), only the first term in Eq. (\ref{eq:psi_2D})
survives, yielding the classical result \cite{Redner:2001a}
\begin{equation}
t_1(\theta) = \frac{R^2}{D_1} (\theta-\epsilon)(2\pi-\epsilon-\theta) .
\end{equation}
The same result is obtained for $a = 0$, since $x^n-1=(1-a/R)^n-1 =
0$.  The limit $a = 0$ is in fact equivalent to the limit $\lambda =
0$ because, after desorption, the particle immediately returns onto
the circle ($a = 0$) as if it was never desorbed ($\lambda=0$).

\subsection{Exact solution}

Iterating the integral equation (\ref{eq:inteq_2d}) shows that the
solution $\psi(\theta)$ writes for $\theta\in[\epsilon,
2\pi-\epsilon]$:
\begin{equation}
\label{eq:psi_2d}
\psi(\theta) = (\theta-\epsilon)(2\pi-\epsilon-\theta) + \sum_{n=1}^\infty d_n \bigl[\cos(n\theta) - \cos(n\epsilon)\bigr] ,
\end{equation}
with the coefficients $d_n$ which satisfy
\begin{equation}
\label{eq:subs_2d}
\begin{split}
& \sum\limits_{n=1}^\infty d_n \bigl[\cos(n\theta) - \cos(n\epsilon)\bigr]  = \Omega \sum\limits_{n=1}^\infty 
\biggl(U_n + \sum\limits_{n'=1}^\infty Q_{n,n'} d_{n'}\biggr)\bigl[\cos(n\theta) - \cos(n\epsilon)\bigr] , \\
\end{split}
\end{equation}
where we introduced
\begin{equation}
\label{eq:U_2d}
\begin{split}
U_n & \equiv \frac{x^n-1}{n^2} 
\int\limits_{\epsilon}^{2\pi-\epsilon} d\theta' \cos (n\theta') (\theta'-\epsilon)(2\pi-\epsilon-\theta')  
 = 4\frac{1-x^n}{n^4}~\xi_n ,\\
\xi_n & \equiv (\pi-\epsilon)\cos(n\epsilon) + \frac{\sin(n\epsilon)}{n}  \hskip 5mm (n = 1,2,...),\\
\end{split}
\end{equation}
and
\begin{equation}
\label{eq:Q_2d}
\begin{split}
Q_{n,n'} & \equiv - \frac{1-x^n}{n^2} I_\epsilon(n,n')  \hskip 5mm (n,n' = 1,2,...) ,\\
\end{split}
\end{equation}
with
\begin{equation}
\label{Ieps}
\begin{split}
I_\epsilon(n,n') & \equiv \int_\epsilon^{2\pi-\epsilon} \cos(n\theta) (\cos(n'\theta)-\cos(n'\epsilon)){\rm d}\theta \\
 & = \left(1-\delta_{n,n'}\right)\left(2\frac{\cos(n'\epsilon)\sin(n\epsilon)}{n}-
\frac{\sin((n'+n)\epsilon)}{n'+n} - \frac{\sin((n'-n)\epsilon)}{n'-n}\right) \\
& + \delta_{n,n'}\left(\pi-\epsilon+\frac{\sin(2n\epsilon)}{2n}\right)  \\
 & = 2\left(1-\delta_{n,n'}\right) \frac{\cos(n\epsilon) \frac{\sin (n'\epsilon)}{n'} - \cos (n'\epsilon) \frac{\sin (n\epsilon)}{n}}{n^2 - n'^2} ~n'^2
 + \delta_{n,n'}\left(\pi-\epsilon+\frac{\sin(2n\epsilon)}{2n}\right) . \\
\end{split}
\end{equation}
Since Eq. (\ref{eq:subs_2d}) should be satisfied for any
$\theta\in[\epsilon, 2\pi-\epsilon]$, one gets ${\bf d} = \Omega (U +
Q{\bf d})$, from which
\begin{equation}
\label{eq:dn_2d}
d_n = \Omega \bigl[(I-\Omega Q)^{-1} U\bigr]_n  \hskip 5mm (n = 1,2,...).
\end{equation}
Since
\begin{equation}
(I-\Omega Q)^{-1} =\sum_{i=0}^\infty (\Omega Q)^{i},
\end{equation}
Eq. (\ref{eq:psi_2d}) with the $d_n$ given by Eq. (\ref{eq:dn_2d}) can
be seen as a series in powers of $\Omega$, whose $n$-th order
coefficient is explicitly written in terms of the $n$-th power of the
matrix $Q$.

Note that the first term in Eq. (\ref{eq:psi_2d}) can also be expended
in a Fourier series
\begin{equation}
\label{eq:theta_Pn}
\sum_{n=1}^\infty e_n \bigl[\cos(n\theta) - \cos(n\epsilon)\bigr] = \begin{cases} 
(\theta-\epsilon)(2\pi-\epsilon-\theta),  ~~~ \epsilon < \theta < 2\pi-\epsilon , \cr
~ \hskip 13mm 0,  \hskip 20mm  \rm{otherwise}, \end{cases}
\end{equation}
where the coefficients $e_n$ are obtained by multiplying this equation
by $\cos m\theta$ and integrating from $0$ to $2\pi$:
\begin{equation}
e_n = - \frac{4}{\pi n^2} \xi_n    \hskip 5mm (n = 1,2,...) .
\end{equation}

Once the $d_n$ determined, the search time $\langle t_1 \rangle$ is
\begin{equation}
\label{eq:t1mean_2d}
\begin{split}
\langle t_1 \rangle & \equiv \frac{1}{2\pi} \int\limits_0^{2\pi} t_1(\theta) d\theta =
\frac{\omega^2 T}{4\pi} \int\limits_{\epsilon}^{2\pi-\epsilon} \psi(\theta) d\theta  
 = \frac{\omega^2 T}{2\pi} \biggl\{ \frac{2}{3}(\pi-\epsilon)^3 - \sum\limits_{n=1}^\infty d_n \xi_n \biggr\} . \\
\end{split}
\end{equation}

\subsection{Approximate solution}

\begin{figure}
\begin{center}
\includegraphics[width=80mm]{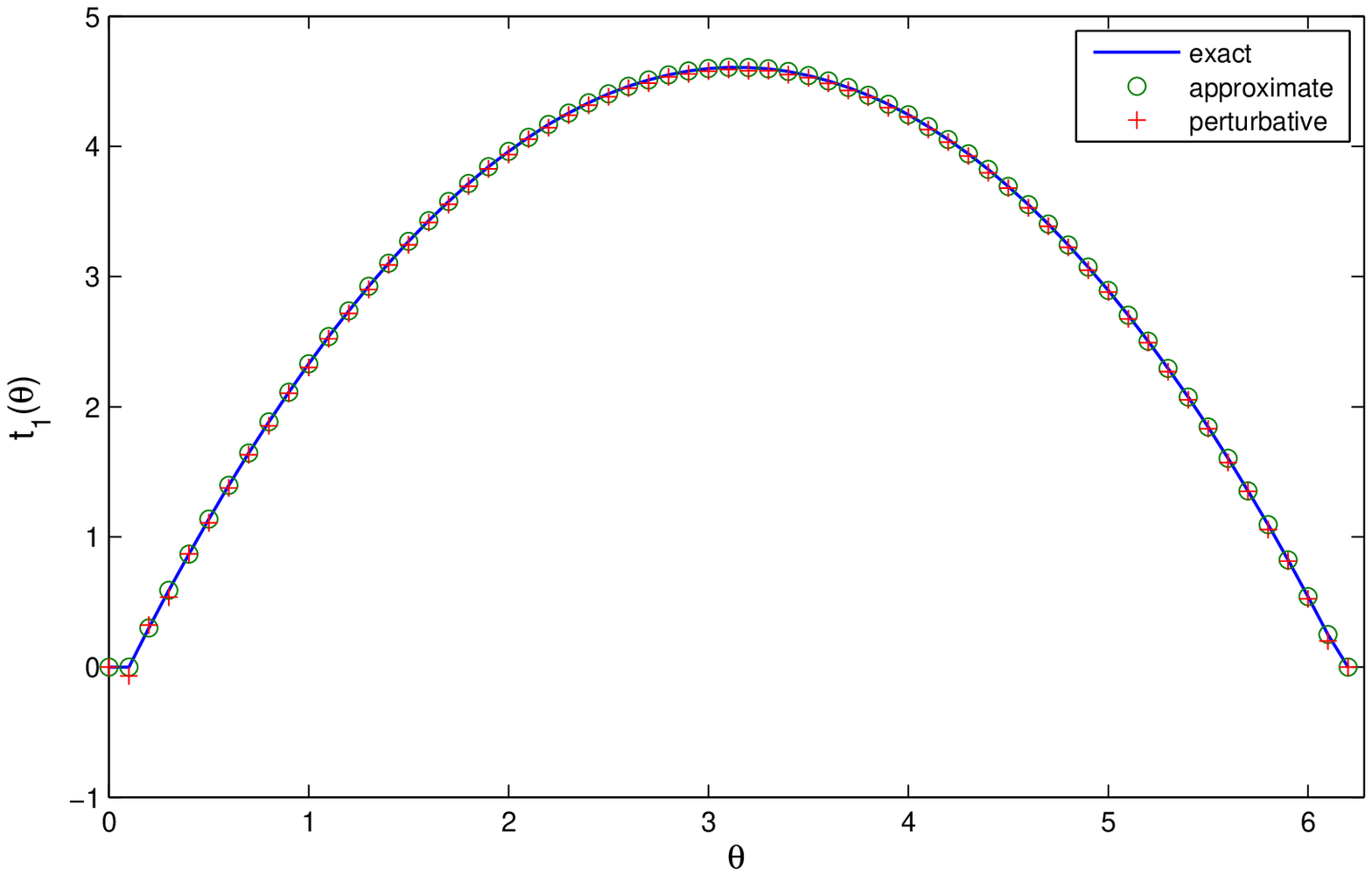}
\includegraphics[width=80mm]{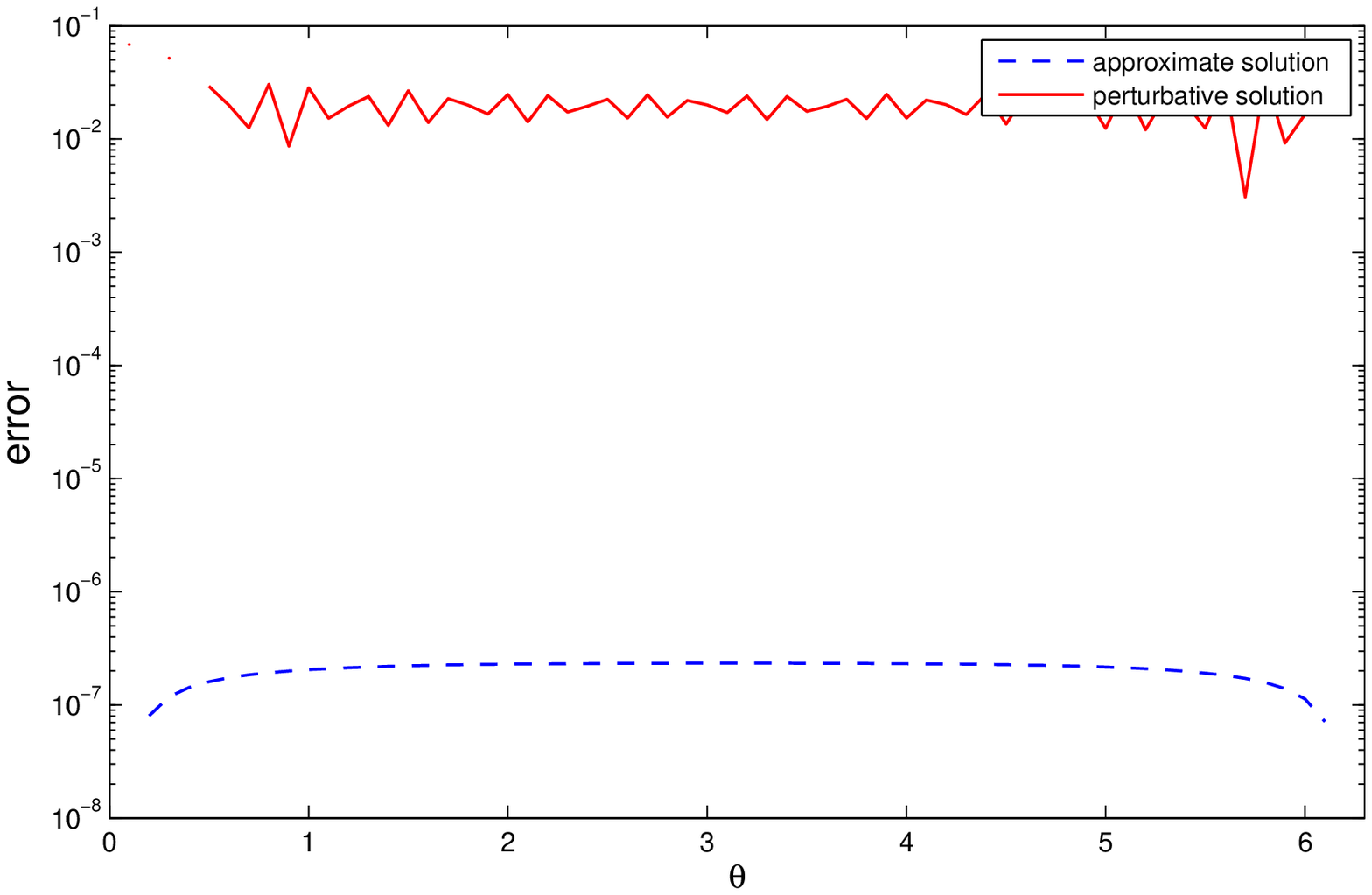}
\includegraphics[width=80mm]{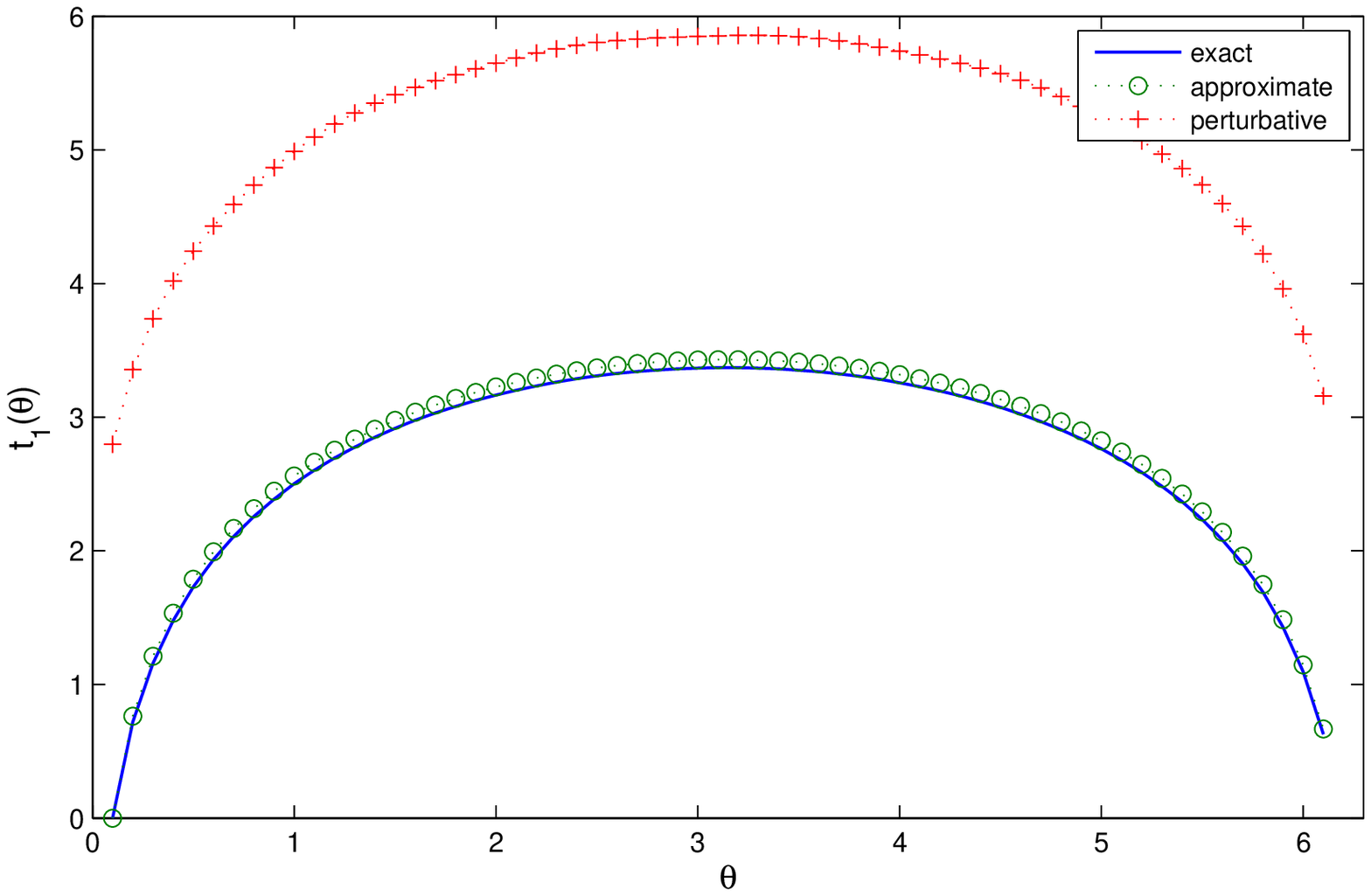}
\includegraphics[width=80mm]{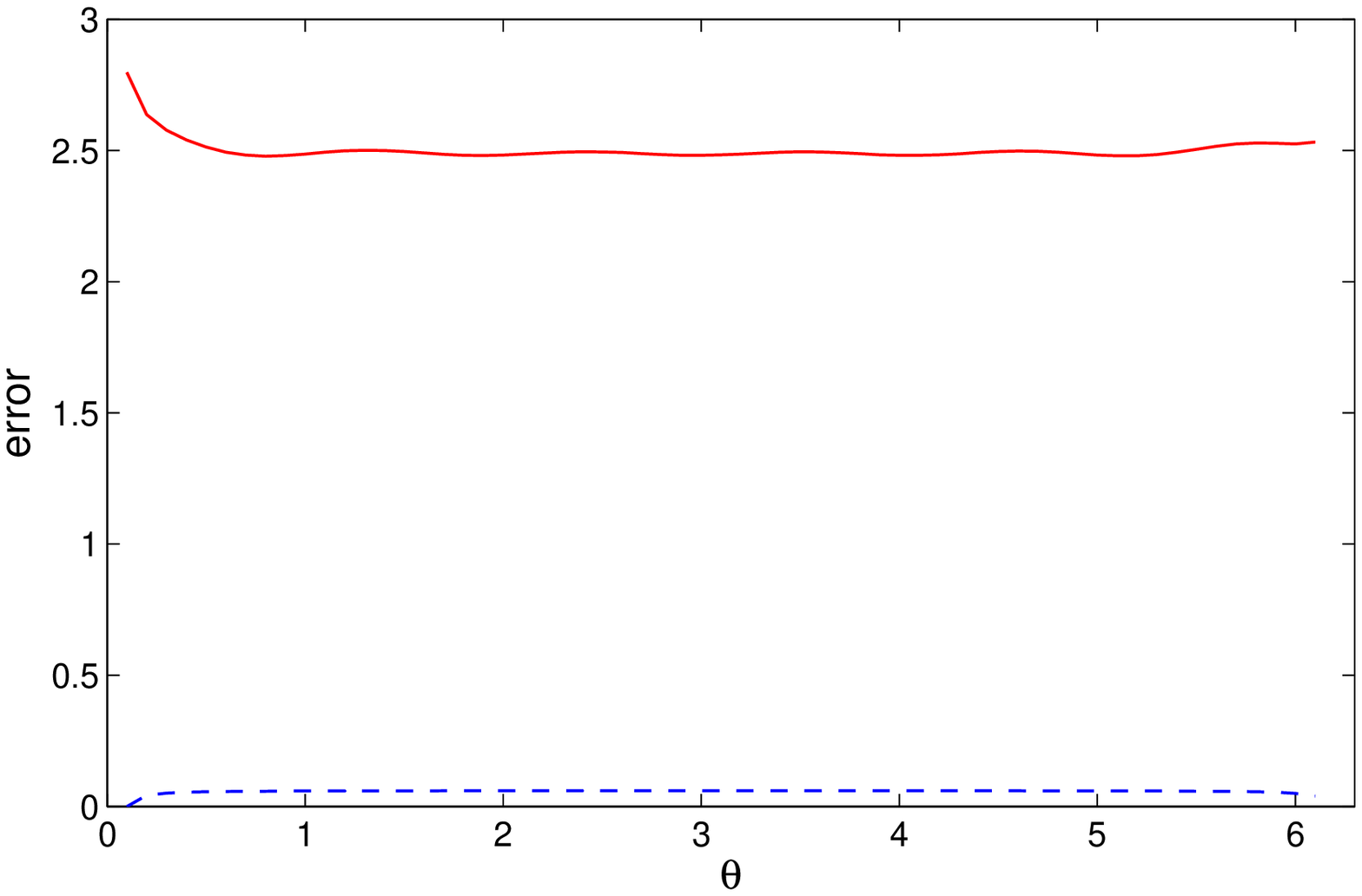}
\includegraphics[width=80mm]{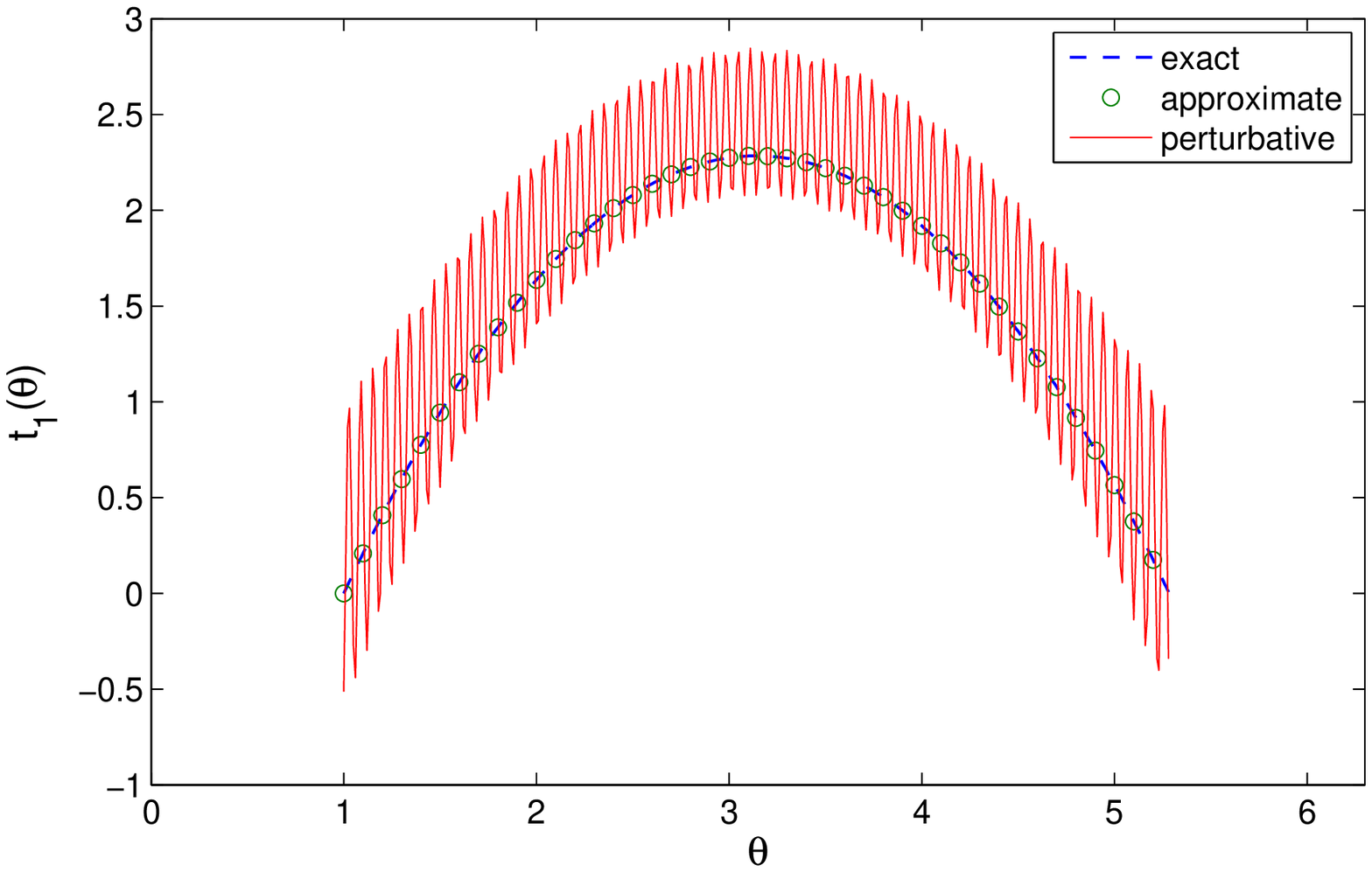}
\includegraphics[width=80mm]{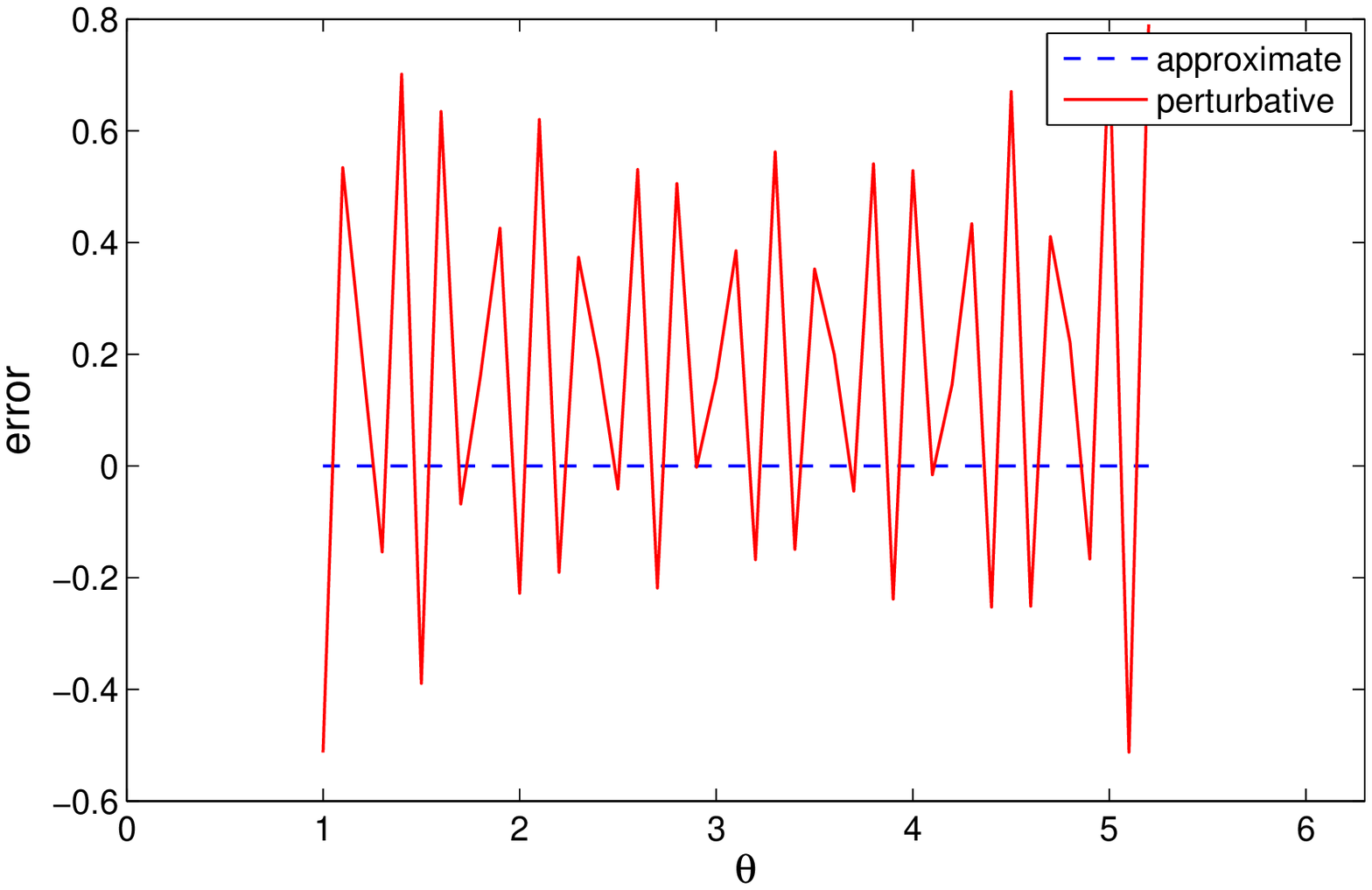}
\includegraphics[width=80mm]{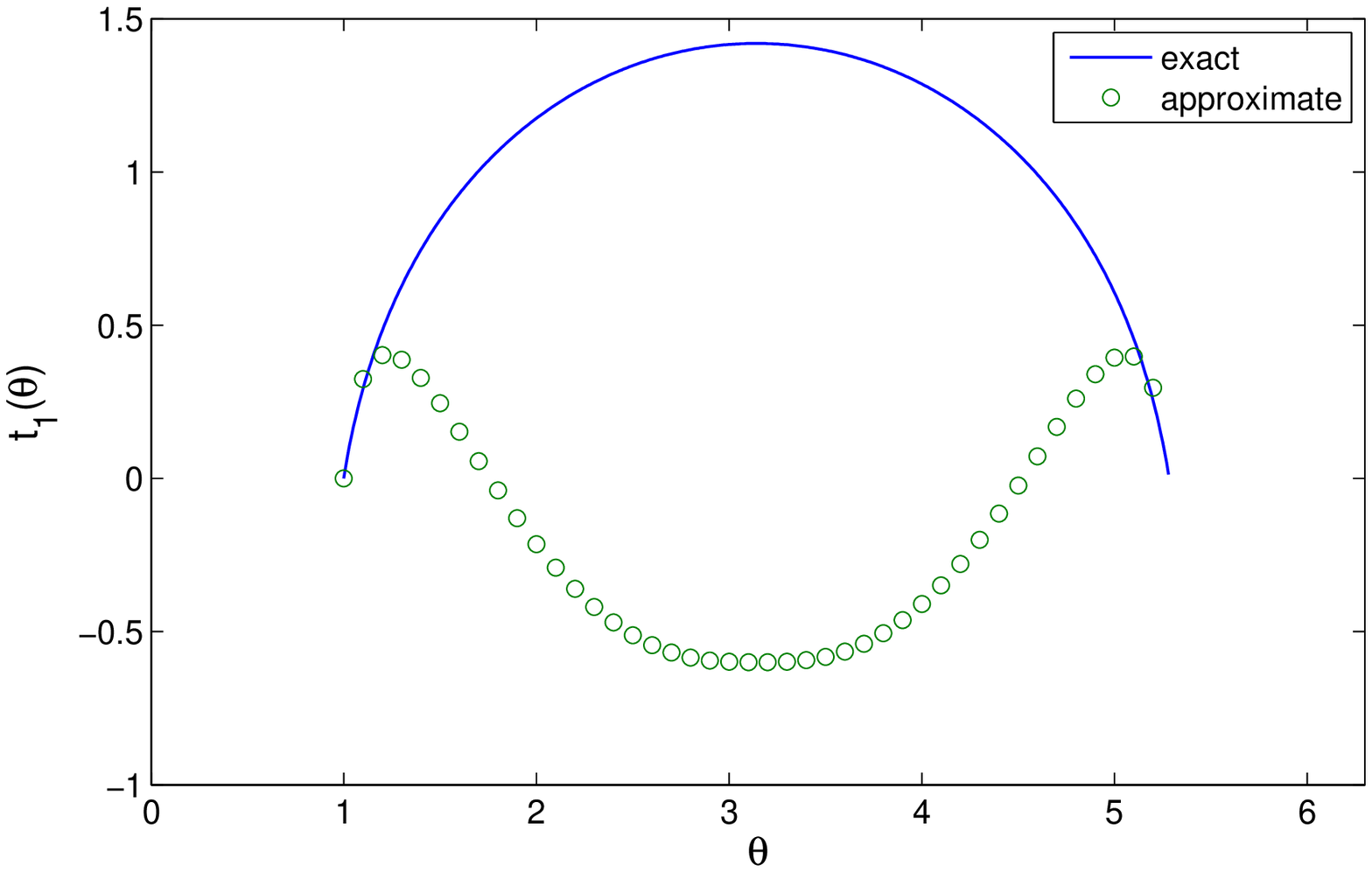}
\end{center}
\caption{
Comparison between three approaches for computing $t_1(\theta)$ in 2D:
the exact solution (\ref{eq:psi_2d}, \ref{eq:dn_2d}), the
approximation (\ref{eq:psi_part_2d}) and the perturbative formula
(\ref{eq:alphanperturbation}), with $D_2 = 1$, $a = 0.01$.  In the
{\bf first row}, the other parameters are: $\epsilon = 0.1$, $\lambda
= 1$, and the series are truncated to $N = 100$.  On the right, the
absolute error between the exact solution and the approximation
(dashed blue curve) and between the exact solution and the
perturbative formula (solid red curve).  The approximation is very
accurate indeed.  In the {\bf second row}, the parameters are:
$\epsilon = 0.1$, $\lambda = 1000$, and the series are truncated to $N
= 100$ for the exact and approximate solutions, and to $N = 1000$ for
the perturbative solution.  One can see that the perturbative solution
is inaccurate for large values of $\lambda$, while the maximal
relative error of the approximate solution is below $2\%$.  In the
{\bf third row}, the parameters are: $\epsilon = 1$, $\lambda = 1$,
and the series are truncated to $N = 100$.  The perturbative solution
is evidently not applicable.  In the {\bf last row}, the parameters
are: $\epsilon = 1$, $\lambda = 1000$, and the series are truncated to
$N = 100$.  In this case, the approximate solution significantly
deviates from the exact one (providing mostly negative values).  The
perturbative solution is completely invalid (not shown). }
\label{2Dtest1}
\end{figure}

\begin{figure}
\begin{center}
\includegraphics[width=80mm]{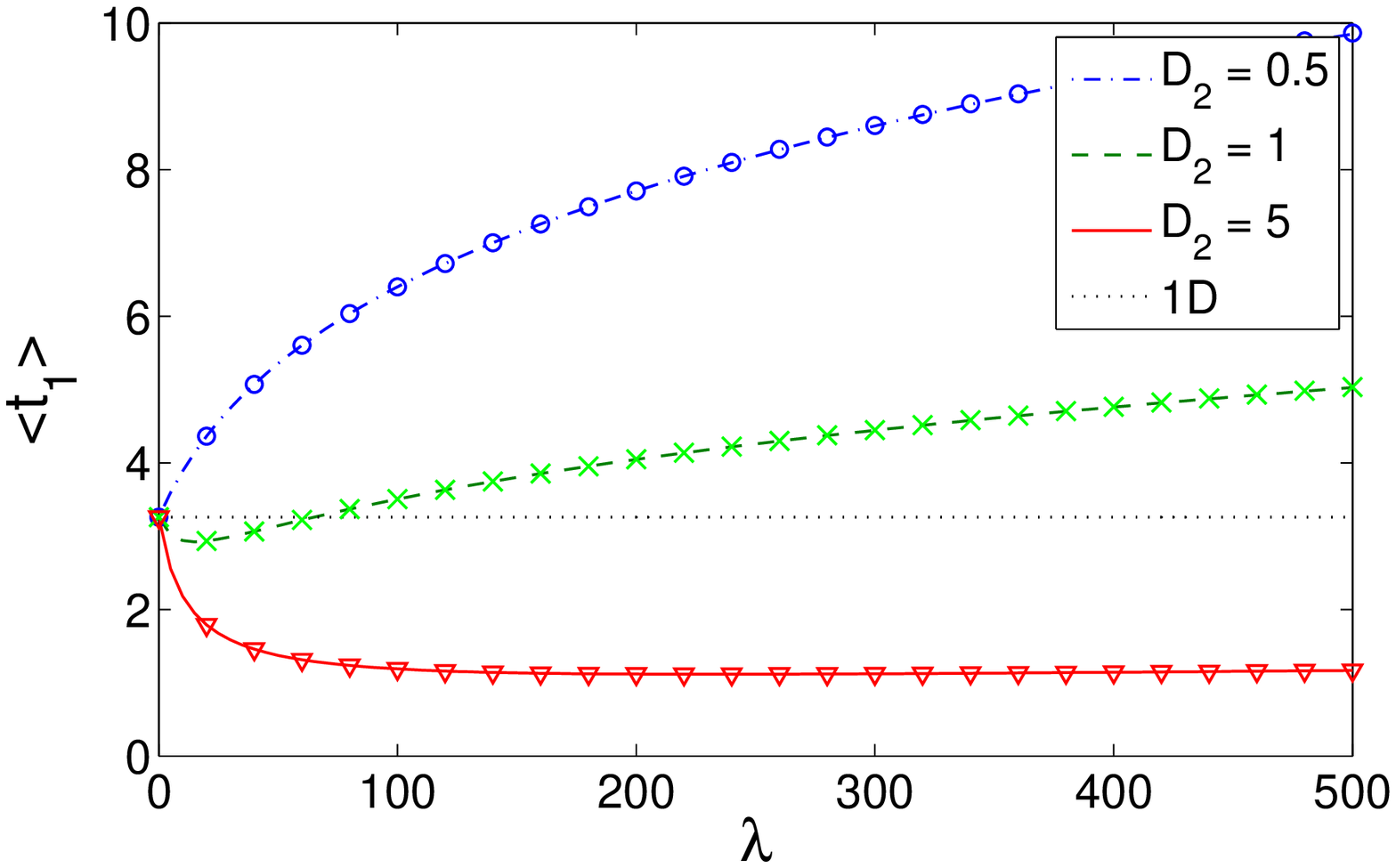}
\includegraphics[width=80mm]{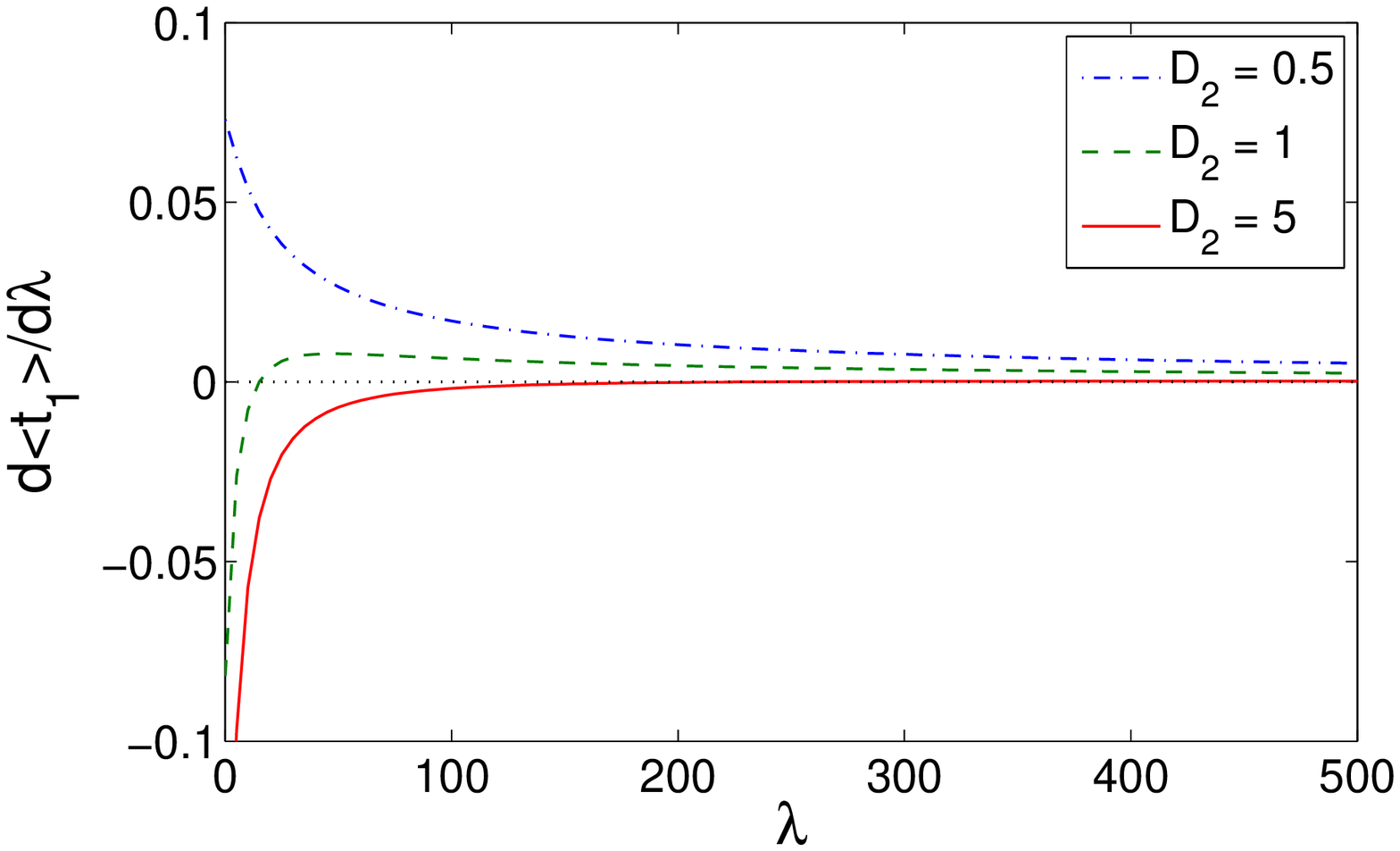}
\end{center}
\caption{
{\bf Left}: In 2D, the mean time $\langle t_1 \rangle$ computed
through Eq. (\ref{eq:dn_2d}, \ref{eq:t1mean_2d}) with $N = 100$ as a
function of the desorption rate $\lambda$ for three values of $D_2$:
$D_2 = 0.5$ (dot-dashed blue line), $D_2 = 1$ (dashed green line), and
$D_2 = 5$ (solid red line).  The other parameters are: $a = 0.1$ and
$\epsilon = 0.01$.  When $D_2 < D_{2,\rm crit} \approx 0.6348...$ (the
first case), $\langle t_1\rangle$ monotonously increases with
$\lambda$ so that there is no optimal value.  In two other cases, $D_2
> D_{2,\rm crit}$, and $\langle t_1 \rangle$ starts first to decrease
with $\lambda$, passes through a minimum (the optimal value) and
monotonously increases.  Symbols show the approximate mean time
computed through Eq. (\ref{eq:betaA_2d}, \ref{eq:t1mean_2d}).  One can
see that the approximation accurate enough even for large values of
$\lambda$.  {\bf Right}: The derivative $\frac{d\langle t_1
\rangle}{d\lambda}$ defined by Eq. (\ref{eq:t1deriv_2d}) for the same
parameters.  }
\label{2Dtest2}
\end{figure}

\begin{figure}
\begin{center}
\includegraphics[width=80mm]{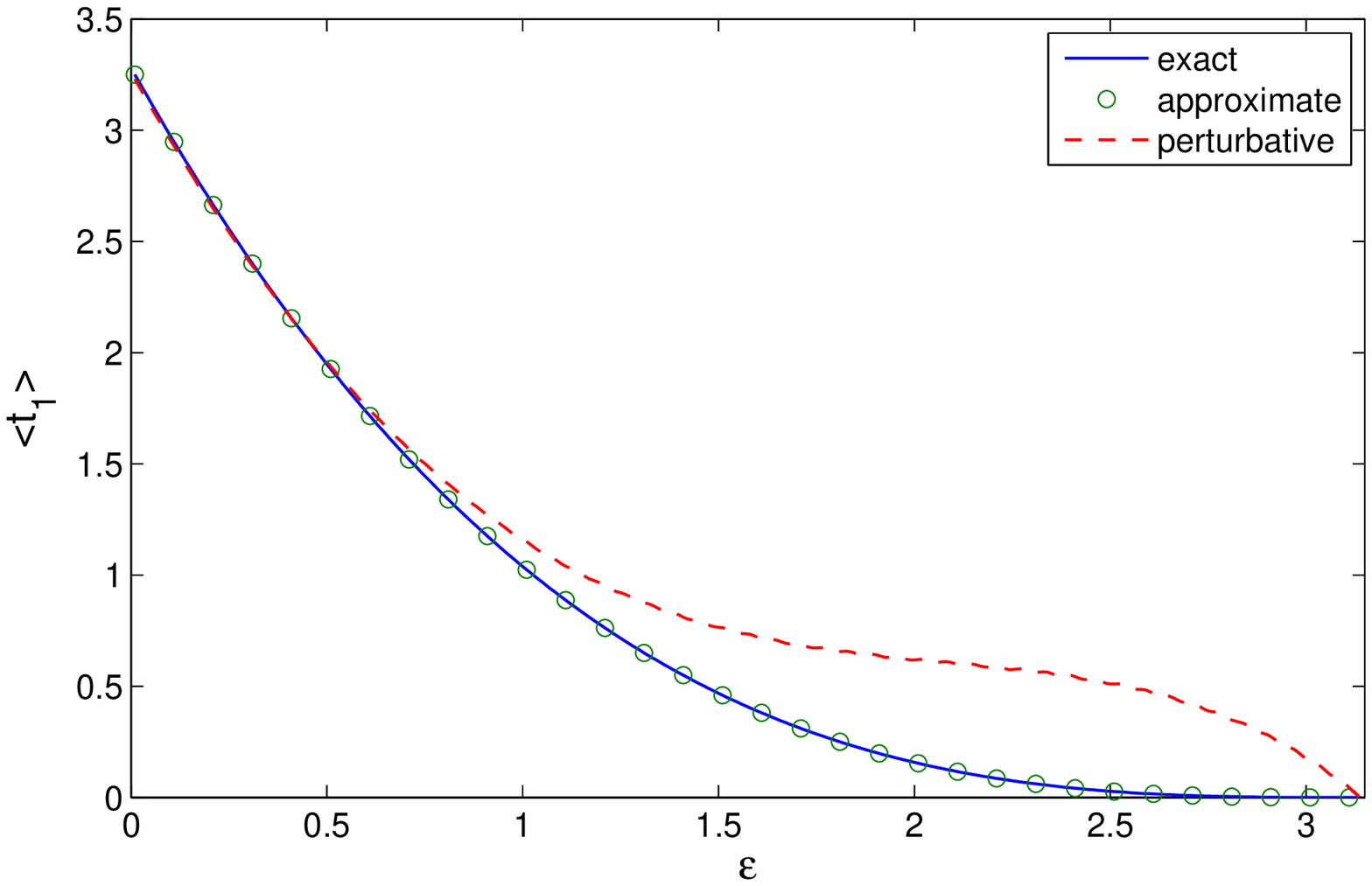}
\includegraphics[width=80mm]{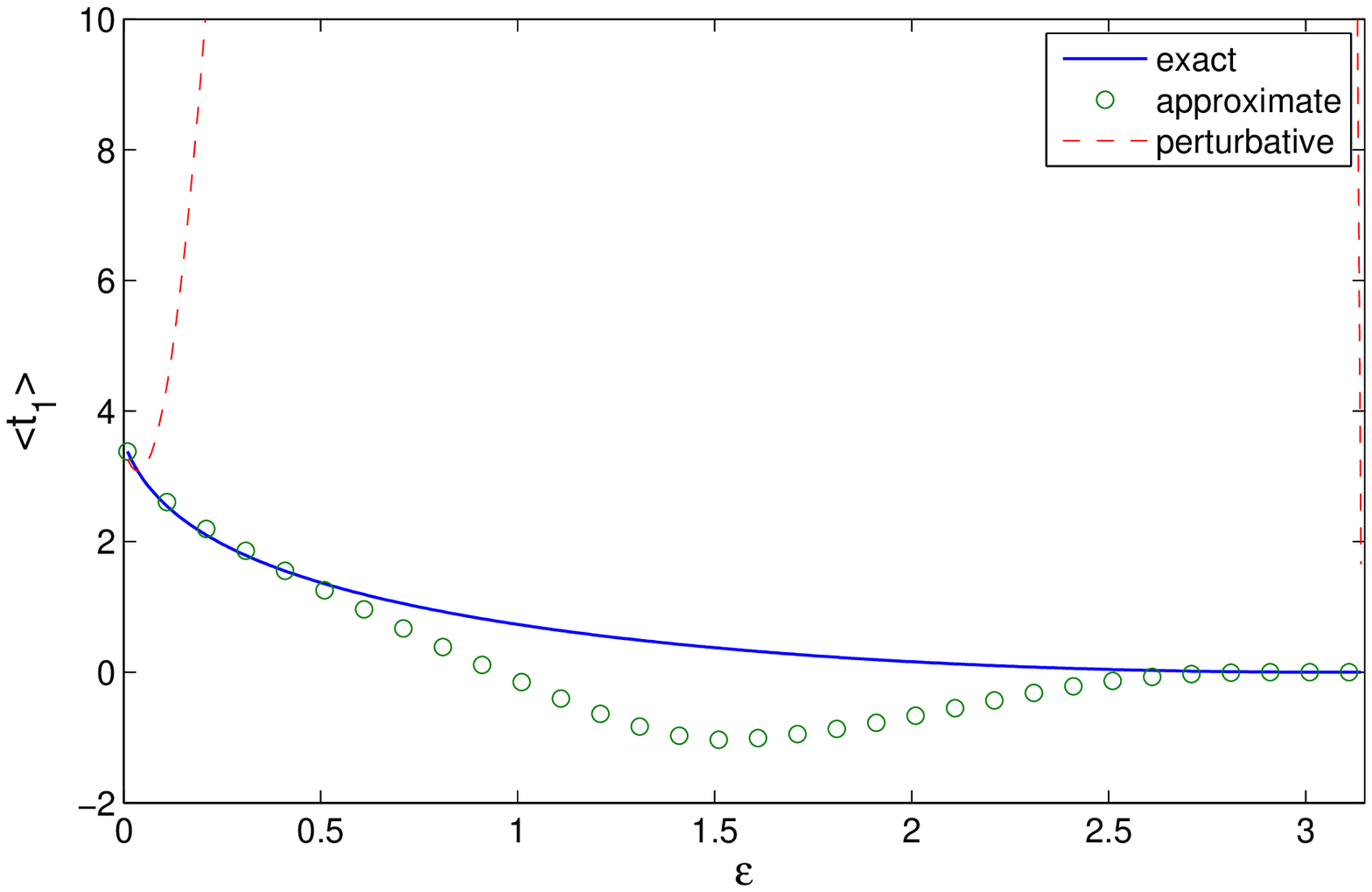}
\end{center}
\caption{
In 2D, the mean time $\langle t_1 \rangle$ as a function of
$\epsilon$, with $D_2 = 1$, $a = 0.01$, and $\lambda = 1$ (left) or
$\lambda = 1000$ (right).  The exact computation through
Eq. (\ref{eq:dn_2d}, \ref{eq:t1mean_2d}) is compared to the
approximation (\ref{eq:betaA_2d}, \ref{eq:t1mean_2d}) and to the
perturbative approach.  In all cases, the series are truncated to $N =
100$.  For small $\lambda$ ($\lambda = 1$), the approximate solution
is very close to the exact one, while the perturbative solution is
relatively close for $\epsilon$ up to $1$.  In turn, for large
$\lambda$ ($\lambda = 1000$), the approximate solution shows
significant deviations for the intermediate values of $\epsilon$,
while the perturbative solution is not applicable at all. }
\label{2Dtest3}
\end{figure}

While the previous expression of $t_1$ is exact, it is not fully
explicit, since it requires either the inversion of the matrix
$I-\Omega Q$ or the calculation of all the powers of $Q$. We give here
an approximation of $(I-\Omega Q)^{-1}$, which in turn provides a
convenient and fully explicit representation of $t_1$.  As shown
numerically (see Figs. \ref{2Dtest1}, \ref{2Dtest2}, \ref{2Dtest3} and
section \ref{numerics} for more details about numerical methods), this
approximation of $t_1$ proves to be in quantitative agreement with the
exact expression for a wide range of parameters.

This approximation relies on the fact that, in the small target size
limit $\epsilon\to 0$, the matrix $Q$ is diagonal, which mirrors the
orthogonality of the $\{\cos(n\theta)\}_n$ on $[0,2\pi]$.  More
precisely, one has from Eqs. (\ref{eq:Q_2d},\ref{Ieps}):
\begin{equation}
Q_{m,n} = \delta_{m,n} Q_{n,n}+{\cal O}(\epsilon^3),
\end{equation}
and keeping only the leading term of this expansion yields
\begin{equation}
\label{eq:betaA_2d}
d_n \approx \Omega (1 - \Omega Q_{n,n})^{-1} U_n ,
\end{equation}
from which we obtain the desired approximation:
\begin{equation}
\label{eq:psi_part_2d}
\psi(\theta)\approx (\theta-\epsilon)(2\pi-\epsilon-\theta) + 4\Omega\sum_{n=1}^\infty
(\cos(n\theta)-\cos(n\epsilon)) \frac{n(\pi-\epsilon)\cos(n\epsilon)+\sin(n\epsilon)}{n^3}
\frac{1-x^n}{n^2+\Omega(1-x^n) I_\epsilon(n,n)}.
\end{equation}
This yields an approximation for the search time:
\begin{equation}
\langle t_1 \rangle \approx \frac{\omega^2 T}{2\pi} \left\{ \frac{2}{3}(\pi-\epsilon)^3 
- 4\Omega \sum\limits_{n=1}^\infty \frac{1-x^n}{n^2} ~
\frac{\bigl((\pi -\epsilon) \cos(n\epsilon) + \frac{\sin (n\epsilon)}{n}\bigr)^2}
{n^2 + \Omega (1-x^n) \bigl(\pi - \epsilon + \frac{\sin(2n\epsilon)}{2n}\bigr)}   \right\} . 
\end{equation}

\subsection{Variations of the search time $\langle t_1\rangle$ with the desorption rate $\lambda$}

In this section, we answer two important questions.  When are bulk
excursions favorable, meaning enabling to reduce the search time (with
respect to the situation with no bulk excursion corresponding to
$\lambda=0$)? If so, is there an optimal value of the desorption rate
$\lambda$ minimizing the search time?

\subsubsection{When are bulk excursions beneficial to the search?}

This question can be investigated by studying the sign of the
derivative $\frac{\partial \langle t_1\rangle}{\partial \lambda}$ at
$\lambda=0$.  The mean search time from Eq. (\ref{eq:t1mean_2d}) can
also be written as
\begin{equation}
\langle t_1\rangle = \frac{R^4}{2\pi^2 D_1^2}\bigl(1 + \lambda \eta\bigr)
\biggl[\frac{2\pi D_1}{3R^2}(\pi - \epsilon)^3 - \lambda \bigl(\xi \cdot (I+\lambda \tilde{Q})^{-1} U\bigr) \biggr],
\end{equation}
where $\tilde{Q} = -Q \frac{R^2}{\pi D_1}$, $\eta = \frac{2aR -
a^2}{4D_2}$.  The derivative of $\langle t_1 \rangle$ with respect to
$\lambda$ is then
\begin{equation}
\label{eq:t1deriv_2d}
\frac{\partial \langle t_1 \rangle}{\partial \lambda} = \frac{R^4 \eta}{2\pi^2 D_1^2}
\biggl[\frac{2\pi D_1}{3R^2}(\pi - \epsilon)^3 - \biggl(\xi \cdot 
\frac{(\eta^{-1} + 2\lambda)I + \lambda^2\tilde{Q}}{(I+\lambda \tilde{Q})^2} U\biggr) \biggr] .
\end{equation}
If the derivative is negative at $\lambda = 0$, i.e.
\begin{equation}
\eta \frac{2\pi D_1}{3R^2} (\pi - \epsilon)^3 < \bigl(\xi \cdot U\bigr) ,
\end{equation}
bulk excursions are beneficial to the search.  This inequality
determines the critical value for the bulk diffusion coefficient
$D_{2,\rm crit}$ (which enters through $\eta$), above which bulk
excursions are beneficial:
\begin{equation}
\label{eq:D2crit_2d}
\frac{D_1}{D_{2,\rm crit}} = \frac{6R^2 (\xi \cdot U)} {\pi (\pi-\epsilon)^3(2aR - a^2)} 
= \frac{24}{\pi (\pi-\epsilon)^3(1-x^2)} \sum\limits_{n=1}^\infty 
\frac{1-x^n}{n^4} \biggl[(\pi -\epsilon) \cos (n\epsilon) + \frac{\sin (n\epsilon)}{n}\biggr]^2.
\end{equation}
Two comments are in order:

(i) Interestingly, this ratio depends only on $a/R$ and $\epsilon$.  In
the limit of $\epsilon\to 0$, one gets
\begin{equation}
\frac{D_1}{D_{2,\rm crit}} \approx \frac{24}{\pi^2(1-x^2)} \sum\limits_{n=1}^\infty \frac{1-x^n}{n^4}.
\end{equation}
Taking next the limit $a/R\to 0$ finally yields:
\begin{equation}
\label{eq:D2crit2_2d}
\frac{D_1}{D_{2,\rm crit}} \approx \frac{12\zeta(3)}{\pi^2} \approx 1.4615,
\end{equation}
where $\zeta$ stands for the Riemann $\zeta$-function.

\begin{figure}
\begin{center}
\includegraphics[width=80mm]{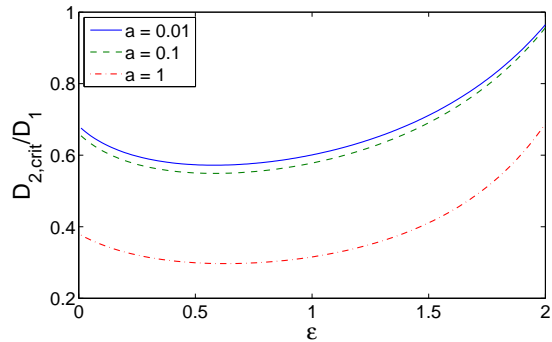}
\end{center}
\caption{
$D_{2,\rm crit}$ as a function of $\epsilon$ is computed from
Eq. (\ref{eq:D2crit_2d}) in 2D for three values of $a/R$: $0.01$,
$0.1$, and $1$.  When $\epsilon$ approaches $\pi$ (the whole surface
becomes absorbing), $D_{2,\rm crit}$ diverges (not shown).  In fact,
in this limit, there is no need for a bulk excursion because the
target will be found immediately by the surface diffusion. }
\label{fig:D2crit_2d}
\end{figure}

(ii) The dependence of the rhs of Eq.(\ref{eq:D2crit_2d}) with
$\epsilon$ is not trivial (Fig. \ref{fig:D2crit_2d}).  Indeed it can
be proved to have a maximum with respect to $\epsilon$, which can be
understood intuitively as follows: in the vicinity of $\epsilon=0$,
increasing $\epsilon$ makes the constraint less stringent since the
target can be reached directly from the bulk; in the opposite limit
$\epsilon\to\pi$, the constraint on $D_1/D_2$ has to tend to $0$ since
the target is found immediately from the surface.  Quantitatively, in
the physical limit $a\to 0$, one finds that, as soon as
$D_2/D_1>(D_{2,\rm crit}/D_1)\approx 0.68...$, bulk excursions can be
beneficial.

\subsubsection{When is there an optimal value of the desorption rate $\lambda$ minimizing the search time?}

If the reaction time $\langle t_1\rangle$ is a decreasing function of
the desorption rate $\lambda$, the bulk excursions are "too favorable",
and the best search strategy is obtained for $\lambda\to\infty$
(purely bulk search).  For the reaction time to be an optimizable
function of $\lambda$, the derivative $\frac{d\langle
t_1\rangle}{d\lambda}$ has to be positive at some $\lambda$.  This
necessary and sufficient condition remains formal and requires
numerical analysis of Eq. (\ref{eq:t1deriv_2d}).  A simple {\it
sufficient} condition can be used instead by demanding that the search
time at zero desorption rate is less than the search time at infinite
desorption rate:
\begin{equation}
\langle t_1(\lambda=0) \rangle < \langle t_1(\lambda\to\infty)\rangle .
\end{equation}

This writes in the physically relevant limit $a\ll R$ (using
the result of \cite{Singer:2006b}):
\begin{eqnarray}
\label{cond2}
\frac{D_1}{D_2}>\frac{(\pi-\epsilon)^3}{3\pi c(\epsilon)}, \;{\rm with}\; c(\epsilon)\equiv 
\frac{1}{\pi\sqrt{2}}\int_0^{\pi -\epsilon} \frac{u\sin(u/2)}{\sqrt{\cos(u)+\cos(\epsilon)}}{\rm d}u.\nonumber\\
\end{eqnarray}

Finally, combining Eqs. (\ref{eq:D2crit_2d}, \ref{cond2}), the search
time is found to be an optimizable function of $\lambda$ in the limit
$a\ll R$ if
\begin{equation}
\label{cond3}
\frac{(\pi-\epsilon)^3}{3\pi c(\epsilon)}<\frac{D_1}{D_2}<\frac{12}{\pi (\pi-\epsilon)^3} \sum\limits_{n=1}^\infty 
\frac{1}{n^3} \biggl[(\pi -\epsilon) \cos (n\epsilon) + \frac{\sin (n\epsilon)}{n}\biggr]^2.
\end{equation}
Knowing that $c(\epsilon)=\ln(2/\epsilon)+{\mathcal O}(\epsilon)$,
Eq. (\ref{cond3}) writes in the small $\epsilon$ limit:
\begin{equation}
\label{eq:optim_region_2d}
\frac{\pi^2}{3\ln(2/\epsilon)}<\frac{D_1}{D_2}<\frac{12\zeta(3)}{\pi^2},
\end{equation}
which summarizes the conditions for the search time to be an
optimizable function of $\lambda$.  This case is illustrated in
Fig. \ref{fig:optimizable}.

\section{3D case}

In this section, the confining domain is a sphere of radius $R$ and
the target is the region on the boundary defined by
$\theta\in[0,\epsilon]$, where $\theta$ is the elevation angle.

\subsection{Basic equations}

The 3D analogs of Eqs. (\ref{eq:t1}, \ref{eq:t2}) read as
\begin{eqnarray}
\frac{D_1}{R^2}\left(\frac{\partial^2t_1}{\partial \theta^2} +\frac{1}{\tan\theta}\frac{\partial t_1}{\partial \theta}\right)+
\lambda [t_2(R-a,\theta)-t_1(\theta)] &=& -1 \;\;{\rm for}\;\theta\in[\epsilon,\pi], \\
D_2\left ( \frac{\partial ^2}{\partial  r^2} +\frac{2}{r} \frac{\partial }{\partial  r} +
\frac{1}{r^2} \frac{\partial ^2}{\partial  \theta^2}\right) t_2(r,\theta) &=& -1. 
\end{eqnarray}
These equations have to be completed by two boundary conditions:
\begin{eqnarray}
\label{eq:adsorption_3d}
 t_2(R,\theta) &=& t_1(\theta), \\
\label{eq:target_3d}
t_1(\theta) &=& 0 \;\;{\rm for} \;\theta\in[0,\epsilon] , 
\end{eqnarray}
which respectively describe the adsorption events and express that the
target is an absorbing region of the sphere.

\subsection{Integral equation for $t_1$}

One can search for a solution in the following form
\begin{equation}
\label{eq:exp3d}
t_2(r,\theta)=\alpha_0-\frac{r^2}{6D_2}+\sum_{n=1}^\infty\alpha_n r^n P_n(\cos \theta),
\end{equation}
where $P_n$ stands for the Legendre polynomial of order $n$.  Using
the orthonormality of Legendre polynomials, the projection of
$t_2(R,\theta)$ on $P_m$ writes
\begin{equation}
\int_0^\pi \sin\theta P_m(\cos\theta)t_2(R,\theta) {\rm d} \theta = 
2\left(\alpha_0-\frac{R^2}{6D_2}\right)\delta_{m,0}+\frac{2\alpha_mR^m}{2m+1}.
\end{equation}
Knowing that 
\begin{equation}
t_2(R,\theta) = \begin{cases} t_1(\theta)  & \text{if $\theta\in[\epsilon,\pi]$,} \\
0 &\text{if $\theta\in[0,\epsilon]$,}
\end{cases}
\end{equation}
the $\alpha_n$ can be written in terms of $t_1(\theta)$ as
\begin{equation}
\label{eq:alpha_n_3D}
\begin{split}
\alpha_0-\frac{R^2}{6D_2} & =\frac{1}{2}\int_\epsilon^\pi\sin\theta\; t_1(\theta){\rm d} \theta , \\
\alpha_nR^n & =\frac{2n+1}{2} \int_\epsilon^\pi\sin\theta\; P_n(\cos\theta)\; t_1(\theta){\rm d} \theta \;\;{\rm if}\;n\ge1. \\
\end{split}
\end{equation}
Taylor expanding the rhs of 
\begin{eqnarray}
\frac{\partial^2t_1}{\partial \theta^2} +\frac{1}{\tan\theta}\frac{\partial t_1}{\partial \theta} = 
-\frac{R^2}{D_1}-\omega^2[t_2(R-a,\theta)-t_2(R,\theta)]
\end{eqnarray}
leads to
\begin{eqnarray}
\frac{\partial^2t_1}{\partial \theta^2} +\frac{1}{\tan\theta}\frac{\partial t_1}{\partial \theta} = 
-\frac{R^2}{D_1}-\omega^2 \sum_{k=1}^\infty \frac{(-a)^k}{k!}\left(\frac{\partial ^k t_2}{\partial r^k}\right)_{R,\theta}.
\end{eqnarray}
Using Eq. (\ref{eq:exp3d}) for $t_2$ yields
\begin{eqnarray}
\frac{\partial^2t_1}{\partial \theta^2} +\frac{1}{\tan\theta}\frac{\partial t_1}{\partial \theta} = 
-\frac{R^2}{D_1}-\omega^2\left(\frac{aR}{3D_2}-\frac{a^2}{6D_2}\right)
-\omega^2 \sum_{k=1}^\infty \frac{(-a)^k}{k!}\sum_{n=k}^\infty \alpha_n n(n-1)\dots (n-k+1) R^{n-k}P_n(\cos \theta).
\end{eqnarray}
Changing the order of summations over $n$ and $k$ and using the
binomial formula and Eq. (\ref{eq:alpha_n_3D}) for $\alpha_n$ finally
give
\begin{eqnarray}
\label{eq:integrodiff_3D}
\frac{\partial^2t_1}{\partial \theta^2} +\frac{1}{\tan\theta}\frac{\partial t_1}{\partial \theta} = 
-\frac{R^2}{D_1}-\omega^2\left(\frac{aR}{3D_2}-\frac{a^2}{6D_2}\right)
-\frac{\omega^2}{2}\sum_{n=1}^\infty (x^n-1) P_n(\cos \theta) (2n+1) \int_\epsilon^\pi \sin\theta' P_n(\cos\theta') t_1(\theta') {\rm d}\theta',
\end{eqnarray}
where, as in previous section, $x\equiv1-\frac{a}{R}$.  This
integro-differential equation for $t_1$ can actually easily be
transformed into an integral equation for $t_1$, by integrating
successively two times. Indeed, multiplying first both members of
Eq. (\ref{eq:integrodiff_3D}) by $\sin\theta$ and integrating between
$\pi$ and $\theta$ gives
\begin{eqnarray}
\label{eq:int3D}
\sin\theta ~t'_1(\theta)&=&\left[\frac{R^2}{D_1}+\omega^2\left(\frac{aR}{3D_2}-\frac{a^2}{6D_2}\right)\right](\cos\theta+1)\nonumber\\
&+&\frac{\omega^2}{2}\sum_{n=1}^\infty (x^n-1) \left(P_{n+1}(\cos\theta)-P_{n-1}(\cos\theta)\right)
\int_\epsilon^\pi \sin\theta' P_n(\cos\theta') t_1(\theta') {\rm d}\theta',
\end{eqnarray}
where we have used 
\begin{equation}
\label{eq:prop_legendre}
\int P_n(x) {\rm d}x = -\frac{1}{n(n+1)}(1-x^2)P'_n(x)=\frac{1}{2n+1}(P_{n+1}(x)-P_{n-1}(x)).
\end{equation}
Dividing Eq. (\ref{eq:int3D}) by $\sin\theta$ and integrating between
$\epsilon$ and $\theta$ finally leads to
\begin{eqnarray*}
t_1(\theta)&=&2\left(\frac{R^2}{D_1}+\omega^2\left(\frac{aR}{3D_2}-\frac{a^2}{6D_2}\right)\right)\ln
\left(\frac{\sin (\theta/2)}{\sin (\epsilon/2)}\right)\nonumber\\
&+&\frac{\omega^2}{2}\sum_{n=1}^\infty (x^n-1) \frac{2n+1}{n(n+1)}
\left(P_{n}(\cos\theta)-P_{n}(\cos\epsilon)\right)
\int_\epsilon^\pi \sin\theta' P_n(\cos\theta') t_1(\theta') {\rm d}\theta',
\end{eqnarray*}
where we have again used Eq. (\ref{eq:prop_legendre}), or equivalently to
\begin{eqnarray}
\label{eq:psi_3D}
\psi(\theta)=\ln\left(\frac{1-\cos\theta}{1-\cos\epsilon}\right)+\Omega\sum_{n=1}^\infty
(x^n-1)\frac{2n+1}{n(n+1)}(P_n(\cos \theta)-P_n(\cos\epsilon))
\int_\epsilon^\pi \sin(\theta')P_n(\cos\theta')\psi(\theta'){\rm d}\theta' ,
\end{eqnarray}
with the following definitions
\begin{equation}
\psi(\theta)\equiv \frac{t_1(\theta)}{\omega^2 T},  \hskip 5mm
T \equiv \frac{1}{\lambda} + \frac{R^2 - (R-a)^2}{6D_2},  \hskip 5mm
\Omega\equiv \frac{\omega^2}{2}
\end{equation}
in this 3D case.

\subsection{Exact solution}

Iterating the integral equation  Eq. (\ref{eq:psi_3D}) shows that the solution $\psi(\theta)$ writes for $\theta\in[\epsilon,\pi]$:
\begin{equation}
\label{eq:psi_3d}
\psi(\theta) = \ln\left(\frac{1-\cos\theta}{1-\cos\epsilon}\right) + 
\sum\limits_{n=1}^\infty d_n \bigl[P_n(\cos\theta) - P_n(\cos\epsilon)\bigr] , 
\end{equation}
with the coefficients $d_n$ which satisfy
\begin{equation}
\label{eq:psi_3D1}
\begin{split}
& \sum\limits_{n=1}^\infty d_n \bigl[P_n(\cos\theta) - P_n(\cos\epsilon)\bigr] 
 = \Omega \sum\limits_{n=1}^\infty \biggl(U_n + \sum\limits_{n'=1}^\infty Q_{n,n'} d_{n'}\biggr)
\bigl[P_n(\cos \theta)-P_n(\cos\epsilon)\bigr] , \\
\end{split}
\end{equation}
where we introduced the new definitions
\begin{equation}
\begin{split}
U_n & \equiv \frac{(x^n-1)(2n+1)}{n(n+1)} \int\limits_\epsilon^\pi {\rm d}\theta' \sin(\theta')P_n(\cos\theta') 
\ln\left(\frac{1-\cos\theta'}{1-\cos\epsilon}\right) 
 = \frac{(1-x^n)(2n+1)}{n^2(n+1)^2}~ \xi_n ,    \\
\xi_n & \equiv \biggl(1 + \frac{n\cos\epsilon}{n+1}\biggr)P_n(\cos\epsilon) + \frac{P_{n-1}(\cos\epsilon)}{n+1} 
\hskip 5mm (n = 1,2,...),   \\
\end{split}
\end{equation}
and
\begin{equation}
Q_{n,n'} \equiv - \frac{(1-x^n)(2n+1)}{n(n+1)} I_\epsilon(n,n')  \hskip 5mm (n,n' = 1,2,...), 
\end{equation}
with
\begin{equation}
\label{eq:I_3d}
I_\epsilon(n,n')\equiv\int\limits_{-1}^{\cos \epsilon}P_n(u)(P_{n'}(u)-P_{n'}(\cos \epsilon)){\rm d}u.
\end{equation}
In Appendix \ref{sec:A_I}, we compute this integral explicitly.

Since Eq. (\ref{eq:psi_3D}) should be satisfied for any $\theta$, one
gets ${\bf d} = \Omega (U + Q {\bf d})$, from which
\begin{equation}
\label{eq:dn_3d}
d_n = \Omega \bigl[(I - \Omega Q)^{-1} U\bigr]_n  \hskip 5mm (n = 1,2,...). 
\end{equation}

As in 2D, using the series expansion of $(I - \Omega Q)^{-1}$,
Eq. (\ref{eq:dn_3d}) can be seen as a series in powers of $\Omega$,
whose $n$-th order coefficient can be explicitly written in terms of
the $n$-th power of the matrix $Q$.

Note that the first term in Eq. (\ref{eq:psi_3d}) can also be
represented as a series
\begin{equation}
\label{eq:log_Pn}
\sum\limits_{n=1}^\infty e_n \bigl[P_n(\cos\theta) - P_n(\cos\epsilon)\bigr] = \begin{cases}
 \ln\left(\frac{1-\cos\theta}{1-\cos\epsilon}\right),   ~~~ \epsilon < \theta < \pi - \epsilon \cr
~ \hskip 8mm  0,  \hskip 13mm  {\rm otherwise}, \end{cases}
\end{equation}
where the coefficients $e_n$ are obtained by multiplying this equation
by $P_n(\cos\theta) \sin\theta$ and integrating from $0$ to $\pi$:
\begin{equation}
e_n = - \frac{2n+1}{2n(n+1)} \xi_n .
\end{equation}

Once the $d_n$ determined, the search time $\langle t_1 \rangle$ can be written as
\begin{equation}
\label{eq:t1mean_3d}
\begin{split}
\langle t_1 \rangle & \equiv  \frac{\omega^2 T}{2}\int\limits_\epsilon^{\pi} d\theta~\sin\theta ~ \psi(\theta) = 
\frac{\omega^2 T}{2} \biggl\{2 \ln\left(\frac{2}{1-\cos\epsilon}\right) - (1+\cos\epsilon) 
 - \sum\limits_{n=1}^\infty d_n \xi_n \biggr\}. \\
\end{split}
\end{equation}

\subsection{Perturbative solution}

The first terms of a perturbative expansion with respect to $\epsilon$
can easily be obtained from the previous exact solution.  At leading
order in $\epsilon$, we have
\begin{equation}
\begin{split}
U_n & = U_n^{(0)} + O(\epsilon) = \frac{2(1-x^n)(2n+1)}{n^2(n+1)^2} + O(\epsilon), \\
Q_{m,n} & = Q_{m,n}^{(0)} + O(\epsilon) = - \frac{2(1-x^n)}{n(n+1)} \delta_{m,n} + O(\epsilon), \\
\end{split}
\end{equation}
from which
\begin{equation}
d_n =\Omega \bigl[(I - \Omega Q^{(0)})^{-1} U^{(0)}\bigr]_n + O(\epsilon) = 
\frac{2\Omega}{n(n+1)} ~ \frac{(1-x^n)(2n+1)}{n(n+1) + 2\Omega (1-x^n)} + O(\epsilon) .
\end{equation}
One finds therefore
\begin{equation}
\label{eq:psi_3d_perturb}
\psi(\theta) = -2\ln\epsilon +2\ln\left(2\sin(\theta/2)\right) - 2\Omega\sum_{n=1}^\infty 
(1-x^n)\frac{2n+1}{n(n+1)}~\frac{1-P_n(\cos\theta)}{n(n+1)+2\Omega(1-x^n)}+O(\epsilon).
\end{equation}
 
Averaging over $\theta$, it finally yields:
\begin{equation}
\langle t_1 \rangle \approx \omega^2 T \biggl\{-2\ln (\epsilon/2) - 1
-2\Omega\sum_{n=1}^\infty \frac{2n+1}{n(n+1)}~\frac{(1-x^n)}{n(n+1)+2\Omega(1-x^n)}+O(\epsilon) \biggr\}.
\end{equation}
This result was given in \cite{Benichou:2010} without derivation.

\subsection{Approximate solution}

\begin{figure}
\begin{center}
\includegraphics[width=80mm]{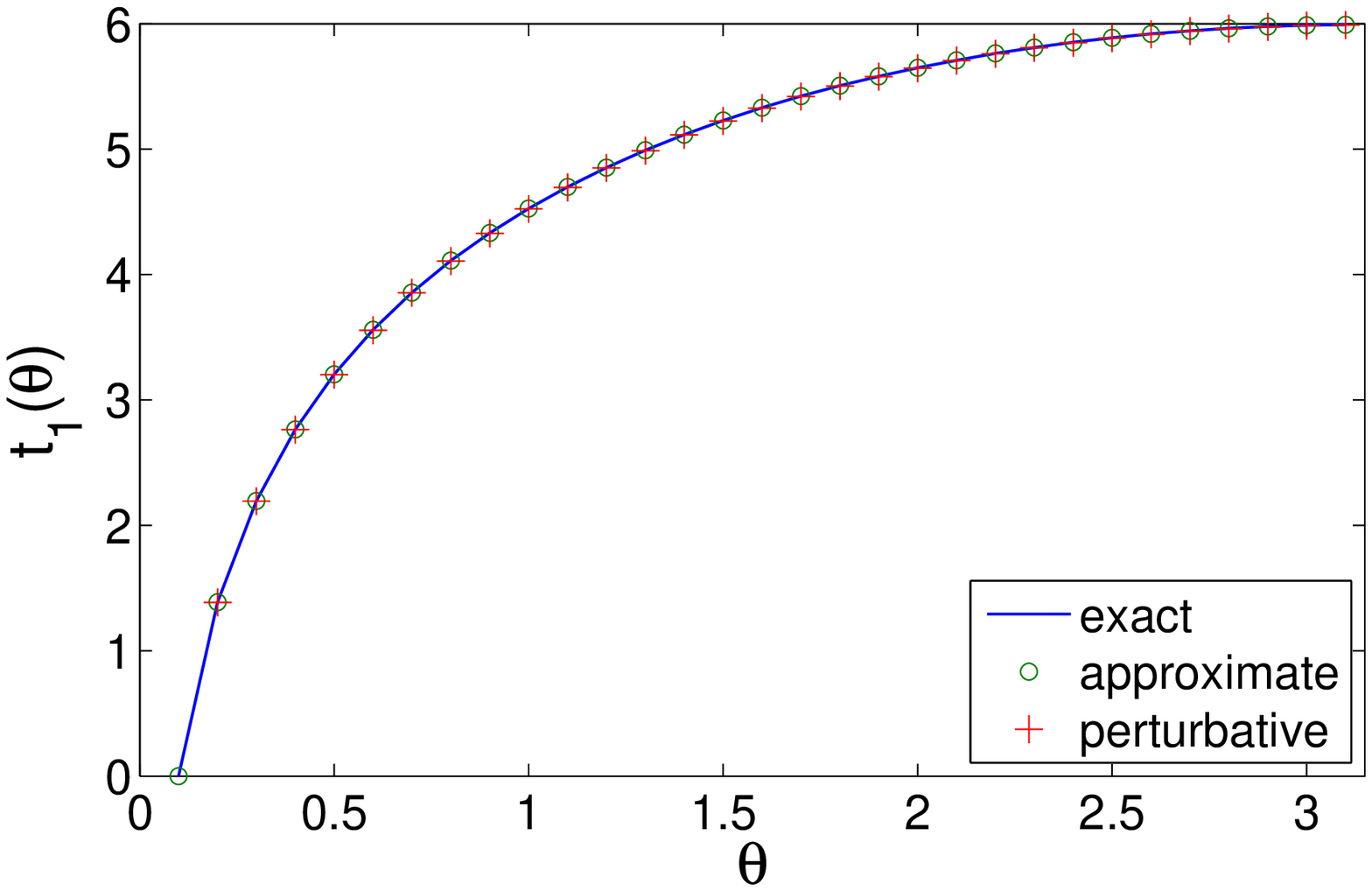}
\includegraphics[width=80mm]{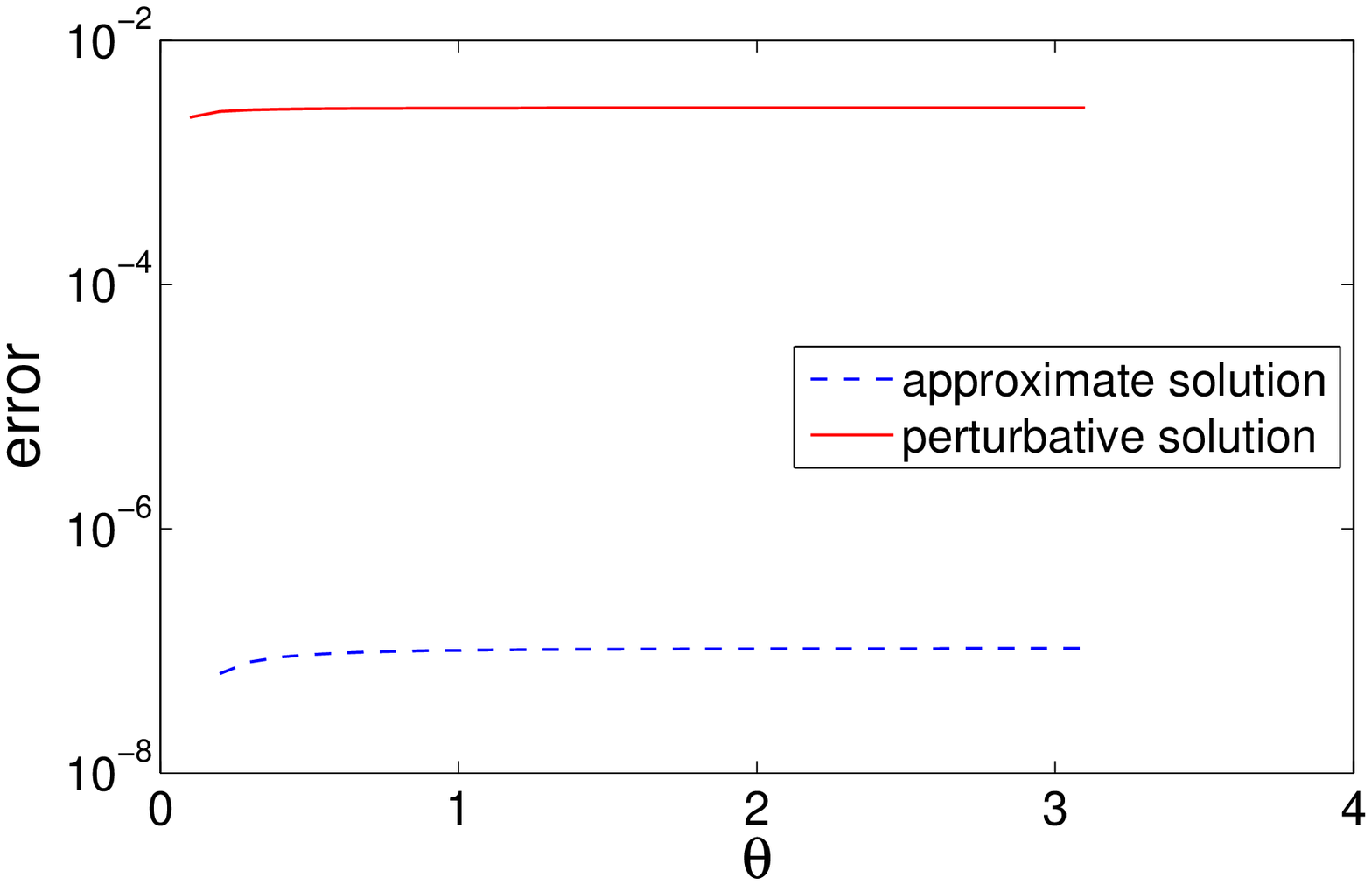}
\includegraphics[width=80mm]{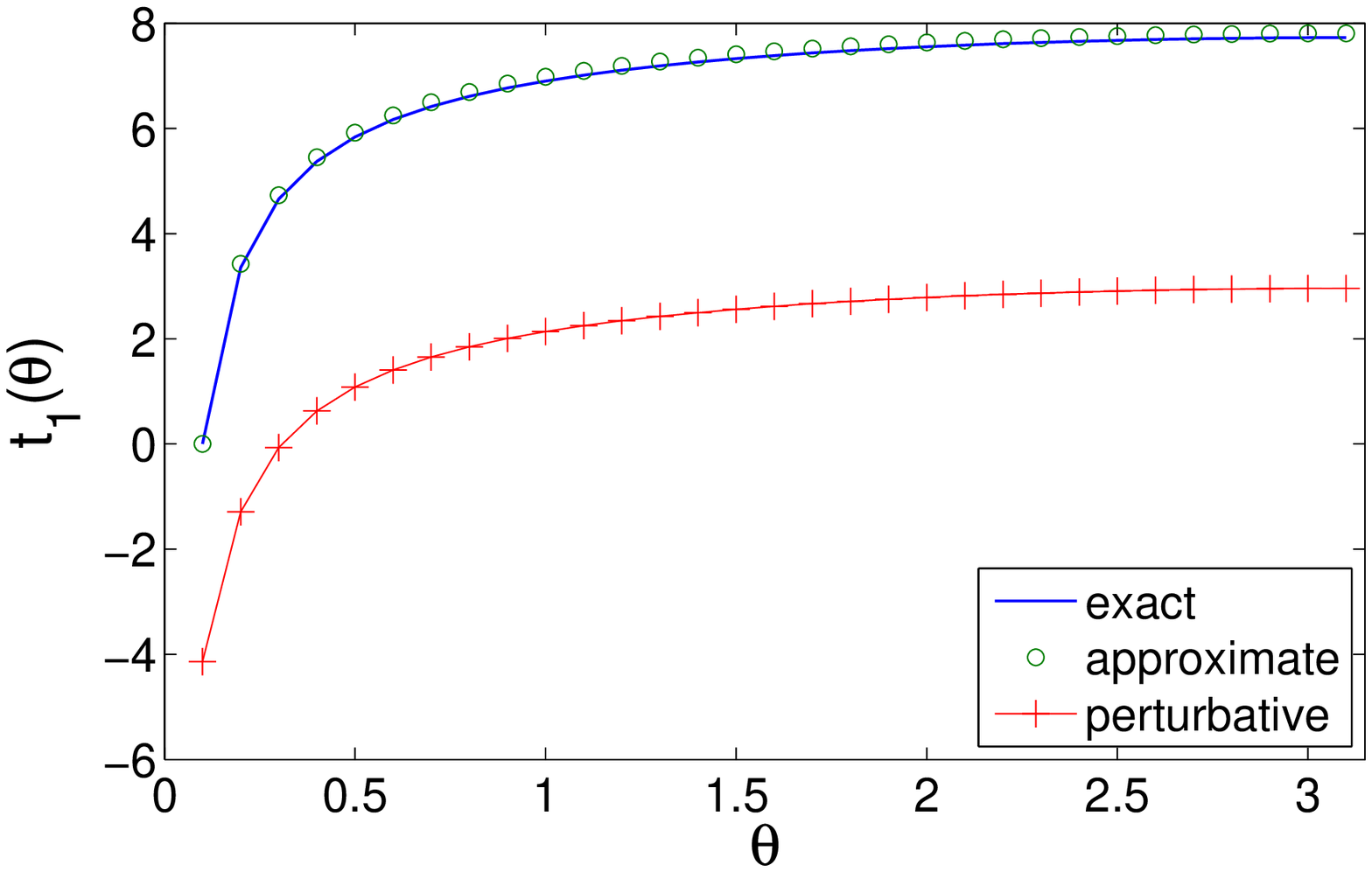}
\includegraphics[width=80mm]{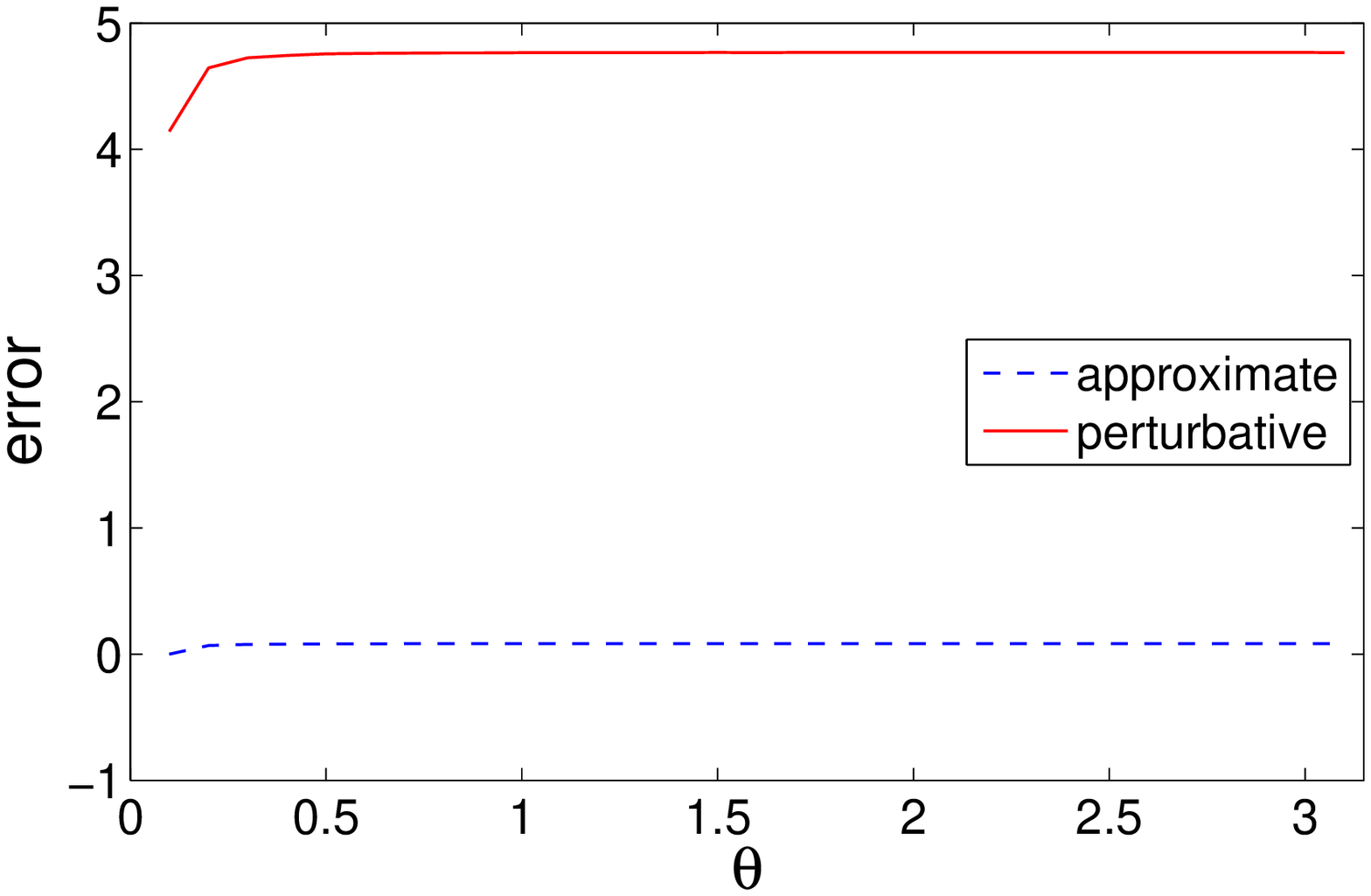}
\includegraphics[width=80mm]{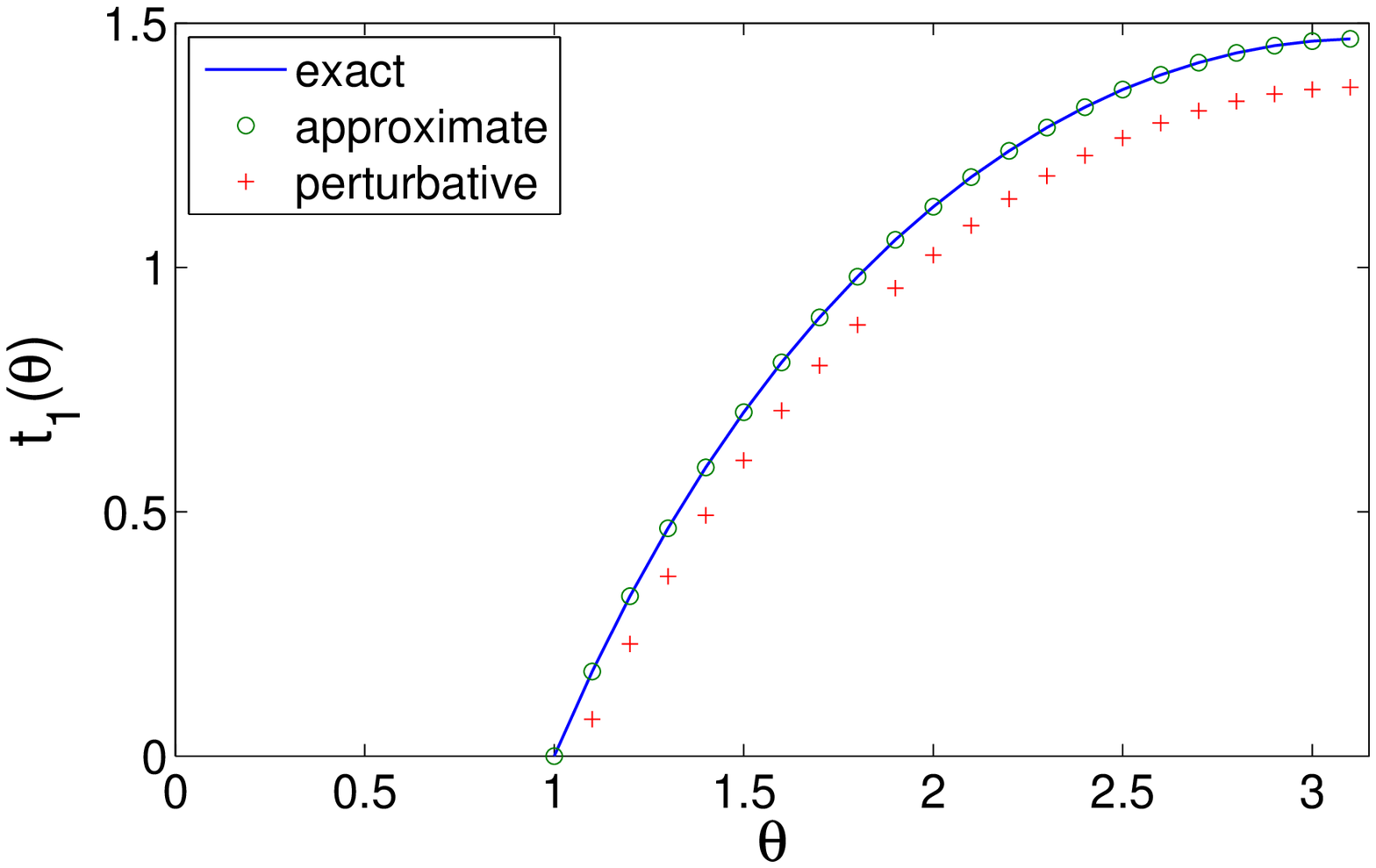}
\includegraphics[width=80mm]{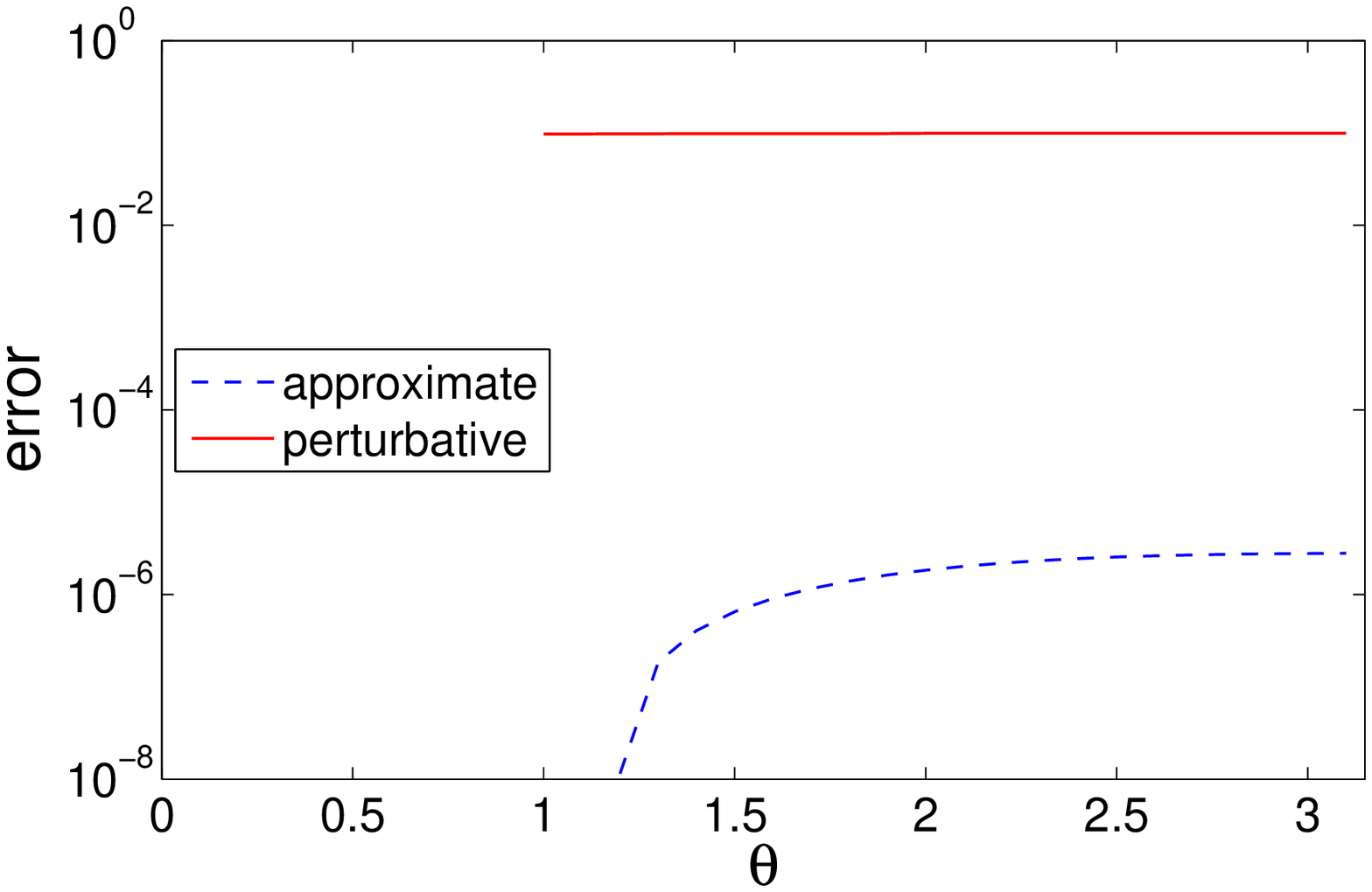}
\includegraphics[width=80mm]{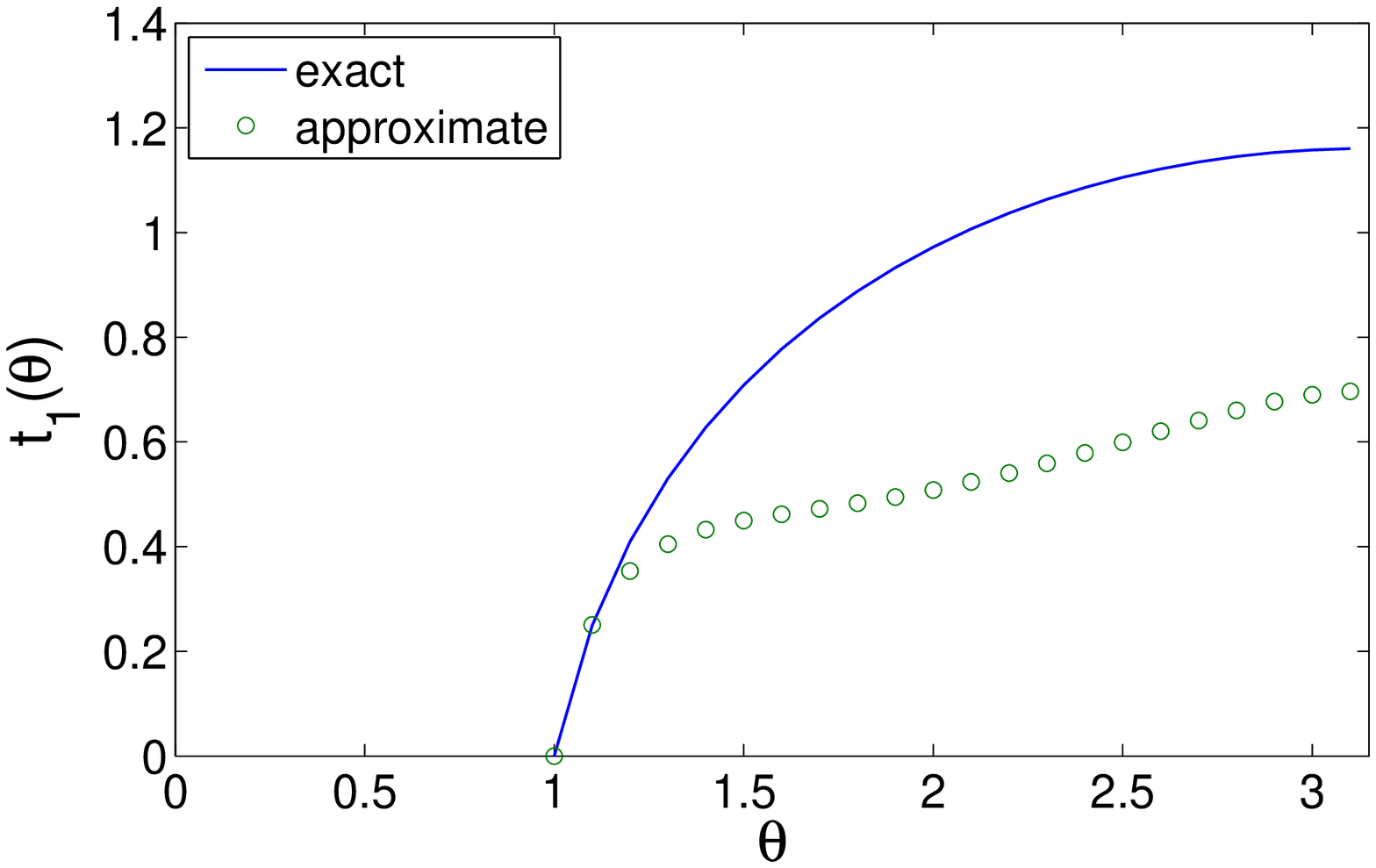}
\end{center}
\caption{
Comparison between three approaches for computing $t_1(\theta)$ in 3D:
the exact solution (\ref{eq:psi_3d}, \ref{eq:dn_3d}), the
approximation (\ref{eq:psi_3d_part}) and the perturbative formula
(\ref{eq:psi_3d_perturb}), with $D_2 = 1$, $a = 0.01$.  In the {\bf
first row}, the other parameters are: $\epsilon = 0.1$, $\lambda = 1$,
and the series are truncated to $N = 100$.  On the right, the absolute
error between the exact solution and the approximation (dashed blue
curve) and between the exact solution and the perturbative formula
(solid red curve).  The approximation is very accurate indeed.  In the
{\bf second row}, the parameters are: $\epsilon = 0.1$, $\lambda =
1000$, and the series are truncated to $N = 100$.  One can see that
the perturbative solution is inaccurate for large values of $\lambda$,
while the maximal relative error of the approximate solution is still
small.  In the {\bf third row}, the parameters are: $\epsilon = 1$,
$\lambda = 1$, and the series are truncated to $N = 100$.  The
perturbative solution is inaccurate as expected for large $\epsilon$.
In the {\bf last row}, the parameters are: $\epsilon = 1$, $\lambda =
1000$, and the series are truncated to $N = 100$.  In this case, the
approximate solution deviates from the exact one for.  The
perturbative solution is negative and not shown. }
\label{3Dtest1}
\end{figure}

\begin{figure}
\begin{center}
\includegraphics[width=80mm]{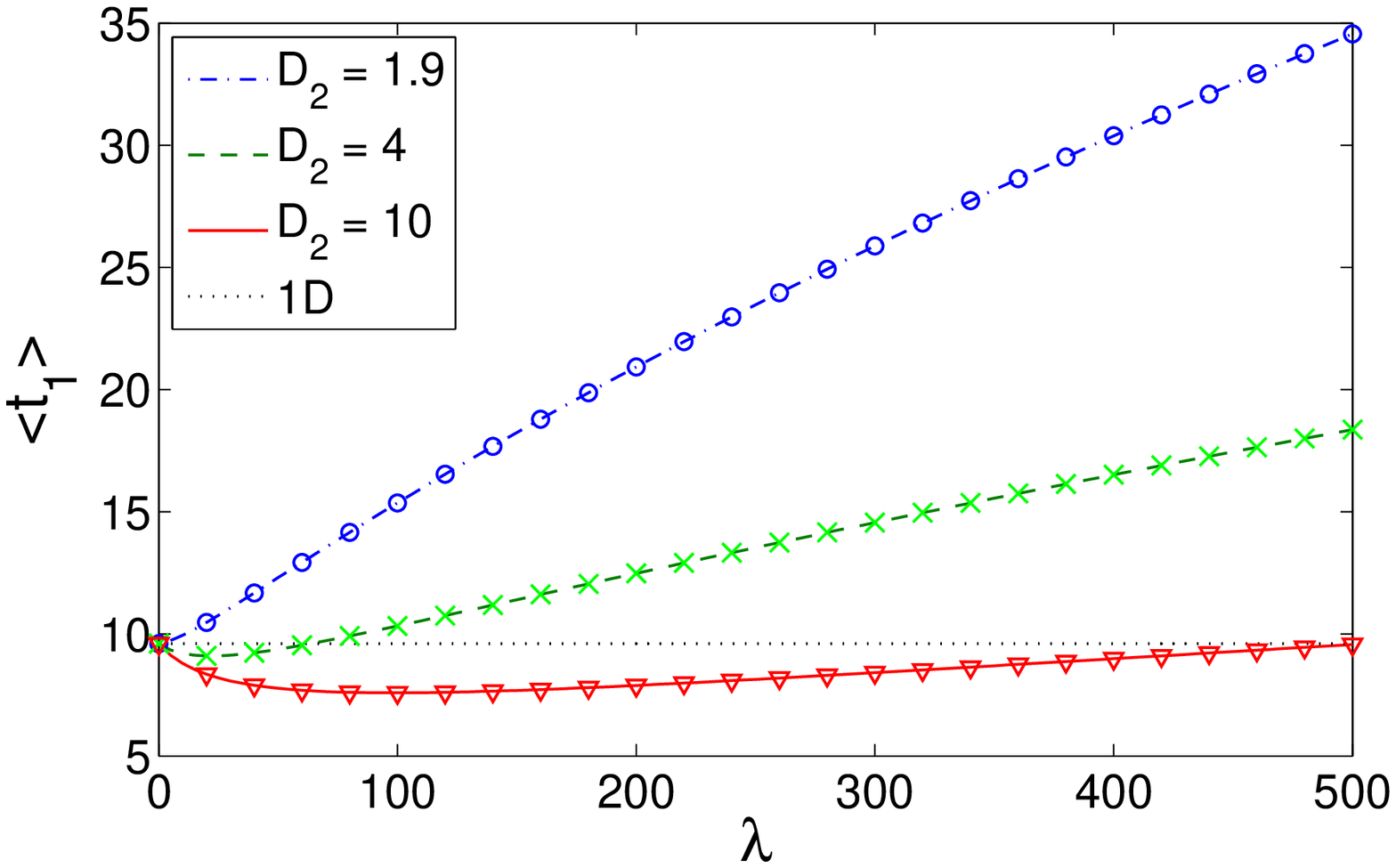}
\includegraphics[width=80mm]{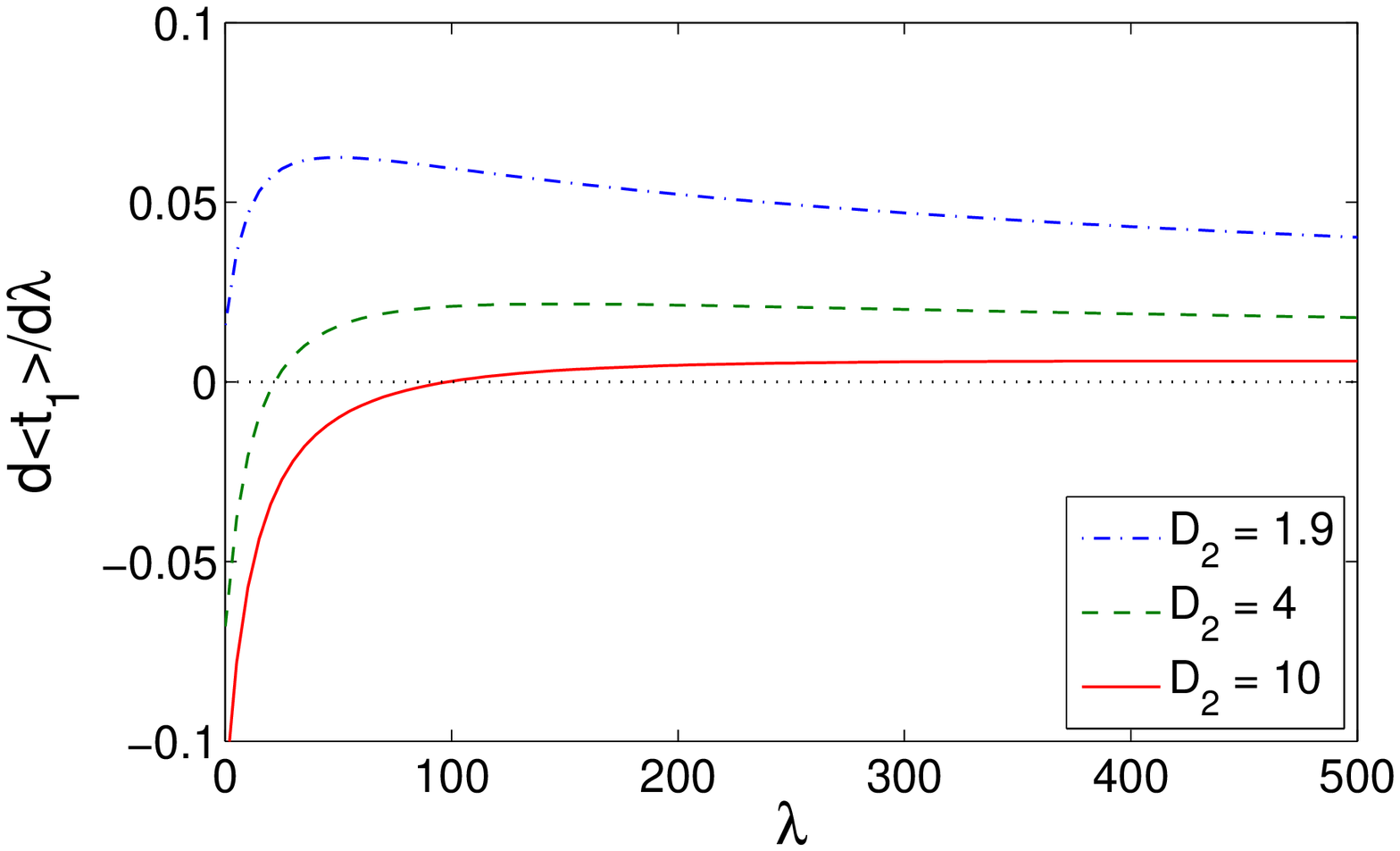}
\end{center}
\caption{
{\bf Left}: In 3D, the mean time $\langle t_1 \rangle$ computed
through Eq. (\ref{eq:dn_3d}, \ref{eq:t1mean_3d}) with $N = 100$ as a
function of the desorption rate $\lambda$ for three values of $D_2$:
$D_2 = 1.9$ (dot-dashed blue line), $D_2 = 4$ (dashed green line), and
$D_2 = 10$ (solid red line).  The other parameters are: $a = 0.1$ and
$\epsilon = 0.01$.  When $D_2 < D_{2,\rm crit} \approx 1.9997....$
(the first case), $\langle t_1 \rangle$ monotonously increases with
$\lambda$ so that there is no optimal value.  In two other cases, $D_2
> D_{2,\rm crit}$, and $\langle t_1 \rangle$ starts first to decrease
with $\lambda$, passes through a minimum (the optimal value) and then
increases.  Symbols show the approximate mean time computed through
Eq. (\ref{eq:psi_3d_av}).  One can see that the approximation accurate
enough even for large values of $\lambda$.  {\bf Right}: The
derivative $\frac{d\langle t_1 \rangle}{d\lambda}$ defined by
Eq. (\ref{eq:t1_deriv_3d}) for the same parameters. }
\label{3Dtest2}
\end{figure}

\begin{figure}
\begin{center}
\includegraphics[width=80mm]{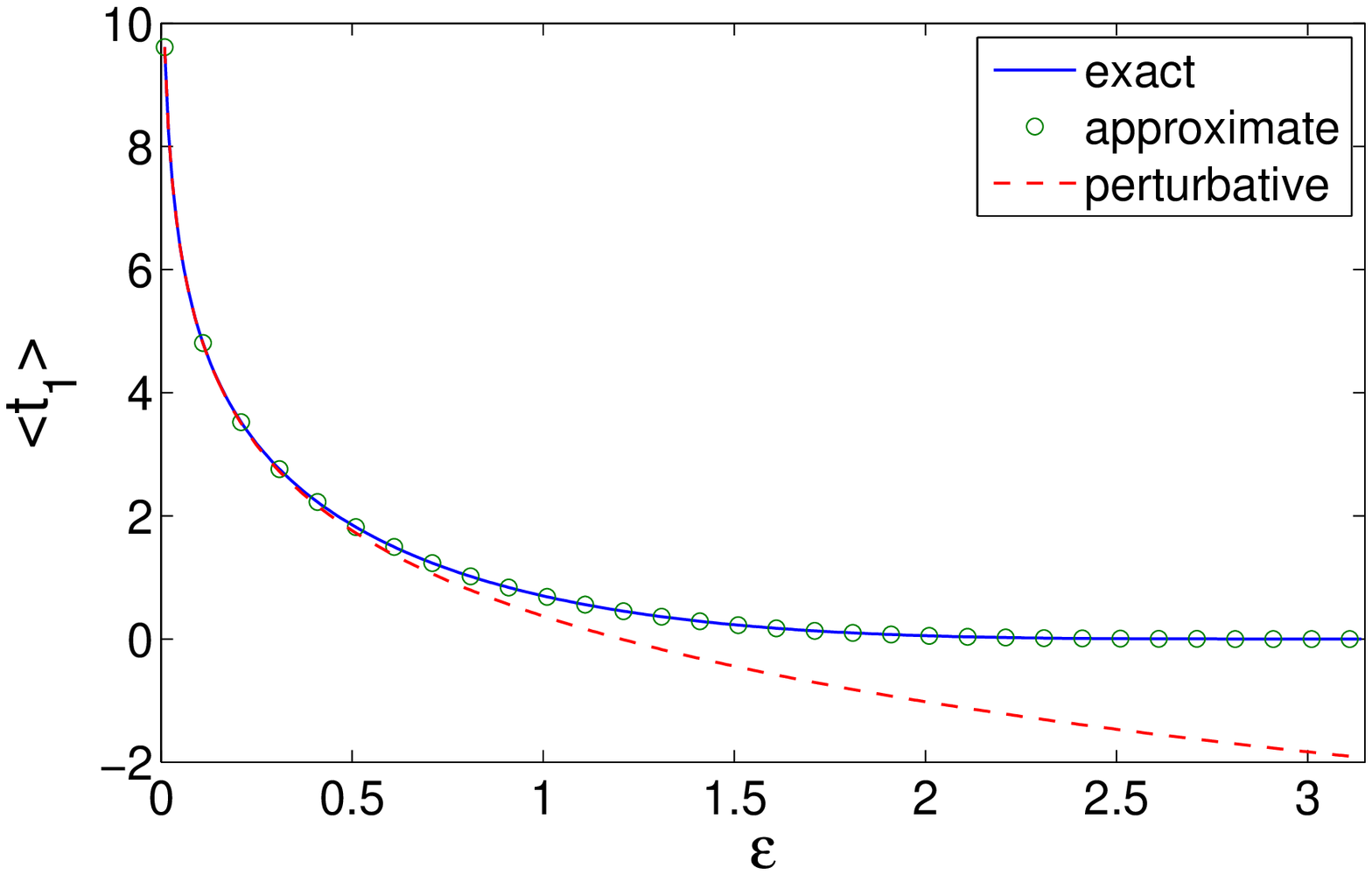}
\includegraphics[width=80mm]{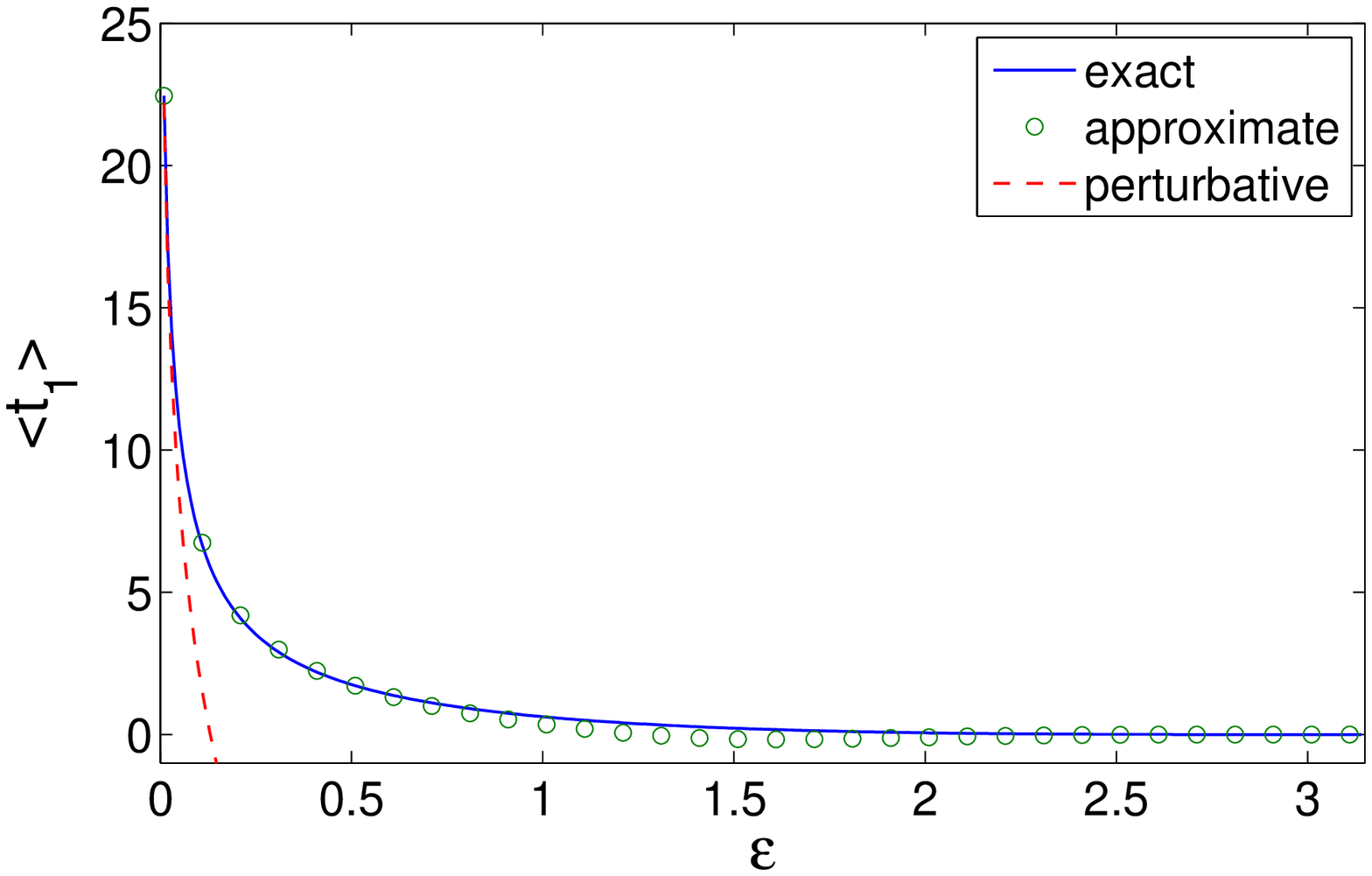}
\end{center}
\caption{
In 3D, the mean time $\langle t_1 \rangle$ as a function of
$\epsilon$, with $D_2 = 1$, $a = 0.01$, and $\lambda = 1$ (left) or
$\lambda = 1000$ (right).  The exact computation through
Eq. (\ref{eq:dn_3d}, \ref{eq:t1mean_3d}) is compared to the
approximation (\ref{eq:psi_3d_av}) and to the perturbative approach.
In all cases, the series are truncated to $N = 100$.  For small
$\lambda$ ($\lambda = 1$), the approximate solution is very close to
the exact one, while the perturbative solution is relatively close for
$\epsilon$ up to $1$.  In turn, for large $\lambda$ ($\lambda =
1000$), the approximate solution shows significant deviations for the
intermediate values of $\epsilon$, while the perturbative solution is
not applicable at all. }
\label{3Dtest3}
\end{figure}

As earlier for the 2D case, an approximate solution can be derived.  As
shown numerically (see Figs. \ref{3Dtest1}, \ref{3Dtest2},
\ref{3Dtest3} and section \ref{numerics} for more details about
numerical methods), this approximation of $t_1$ proves to be in
quantitative agreement with the exact expression for a wide range of
parameters.

This approximation relies on the fact that, in the small target size
limit $\epsilon\to 0$, the matrix $Q$ is diagonal, which in turn
mirrors the orthogonality of $\{P_n(\cos\theta)\}_n$ on $[0,\pi]$.
More precisely, one has
\begin{equation}
Q_{m,n} = \delta_{m,n} Q_{n,n}+{\cal O}(\epsilon^4),
\end{equation}
and keeping only the leading term of this expansion yields
\begin{equation}
d_n \approx \Omega (1 - \Omega Q_{n,n})^{-1} U_n ,
\end{equation}
from which
\begin{equation}
\label{eq:psi_3d_part}
\psi(\theta)\approx \ln\left(\frac{1-\cos \theta}{1-\cos \epsilon}\right) + \Omega \sum_{n=1}^\infty 
(1-x^n) \frac{2n+1}{n(n+1)}(P_n(\cos \theta)-P_n(\cos \epsilon))
\frac{\bigl(1 + \frac{n\cos\epsilon}{n+1}\bigr)P_n(\cos\epsilon) + \frac{P_{n-1}(\cos\epsilon)}{n+1}}
{n(n+1) + \Omega(1-x^n)(2n+1)I_\epsilon(n,n)}.
\end{equation}

The mean time $\langle t_1 \rangle$ is then approximated as
\begin{equation}
\label{eq:psi_3d_av}
\begin{split}
\langle t_1 \rangle & \approx \omega^2 T \biggl\{\ln\left(\frac{2}{1-\cos\epsilon}\right)
- \frac{1+\cos\epsilon}{2} \\
& - \frac{\Omega}{2} \sum\limits_{n=1}^\infty 
\frac{(1-x^n)(2n+1)}{n(n+1)} \frac{\bigl[\bigl(1 + \frac{n\cos\epsilon}{n+1}\bigr)P_n(\cos\epsilon) + 
\frac{P_{n-1}(\cos\epsilon)}{n+1}\bigr]^2}{n(n+1) + \Omega (1-x^n)(2n+1) I_\epsilon(n,n)}\biggr]\biggr\}. \\
\end{split}
\end{equation}

\subsection{Variations of the search time $\langle t_1 \rangle$ with the desorption rate $\lambda$}

We investigate here as in the 2D case the dependence of $\langle t_1
\rangle$ on $\lambda$.

\subsubsection{When are bulk excursions beneficial to the search?}

The sign of $\frac{\partial \langle t_1 \rangle}{\partial \lambda}$ at
$\lambda=0$ is conveniently studied by rewriting
Eq. (\ref{eq:t1mean_3d}) as
\begin{equation}
\langle t_1 \rangle = \frac{R^4}{4D_1^2}\bigl(1 + \lambda \eta\bigr)
\biggl\{\frac{4D_1}{R^2} \biggl[\ln\left(\frac{2}{1-\cos\epsilon}\right) - \frac{1+\cos\epsilon}{2}\biggr]
- \lambda \bigl(\xi \cdot (I+\lambda \tilde{Q})^{-1} U\bigr) \biggr\} ,
\end{equation}
where $\tilde{Q} = -Q \frac{R^2}{2D_1}$ and $\eta = \frac{2aR -
a^2}{6D_2}$.  The derivative of $\langle t_1 \rangle$ with respect to
$\lambda$ is then
\begin{equation}
\label{eq:t1_deriv_3d}
\frac{\partial \langle t_1 \rangle}{\partial \lambda} = \frac{R^4 \eta}{4D_1^2}
\biggl\{\biggl[\ln\left(\frac{2}{1-\cos\epsilon}\right) - \frac{1+\cos\epsilon}{2}\biggr]
- \biggl(\xi \cdot \frac{(\eta^{-1} + 2\lambda)I + \lambda^2 \tilde{Q}}{(I+\lambda\tilde{Q})^2} U\biggr)\biggr\} .
\end{equation}

If the above derivative is negative at $\lambda = 0$, i.e.
\begin{equation}
\frac{4D_1}{R^2}~ \biggl(\ln\left(\frac{2}{1-\cos\epsilon}\right) - \frac{1+\cos\epsilon}{2}\biggr) < \frac{(\xi \cdot U)}{\eta} ,
\end{equation}
bulk excursions are beneficial to the search.  This inequality
determines the critical value for the bulk diffusion coefficient
$D_{2,\rm crit}$ (which enters through $\eta$):
\begin{equation}
\label{eq:D2crit_3d}
\begin{split}
\frac{D_{2,\rm crit}}{D_1} & = \biggl(\ln\left(\frac{2}{1-\cos\epsilon}\right) - \frac{1+\cos\epsilon}{2}\biggr) 
\frac{2(2aR-a^2)}{3R^2 (\xi \cdot U)} \\
& = \biggl(\ln\left(\frac{2}{1-\cos\epsilon}\right) - \frac{1+\cos\epsilon}{2}\biggr) \frac{2(1-x^2)}{3} \\
& \times \biggl(\sum\limits_{n=1}^\infty \frac{(1-x^n)(2n+1)}{n^2(n+1)^4} 
\bigl[(n+1 + n\cos\epsilon) P_n(\cos\epsilon) + P_{n-1}(\cos\epsilon)\bigr]^2 \biggr)^{-1} . \\
\end{split}
\end{equation}
In the limit of $\epsilon\to 0$, one gets
\begin{equation}
\frac{D_{2,\rm crit}}{D_1} \approx \frac{(2\ln(2/\epsilon) - 1)(1-x^2)}{6} 
\biggl(\sum\limits_{n=1}^\infty \frac{(1-x^n)(2n+1)}{n^2(n+1)^2} \biggr)^{-1} + O(\epsilon).
\end{equation}
In the physically relevant limit $a\ll R$, one has
\begin{equation}
\label{eq:D2crit2_3d}
\frac{D_{2,\rm crit}}{D_1} \approx \frac{(2\ln(2/\epsilon) - 1)}{3} 
\biggl(\sum\limits_{n=1}^\infty \frac{2n+1}{n(n+1)^2} \biggr)^{-1} + O(\epsilon) =
\frac{2(2\ln(2/\epsilon) - 1)}{\pi^2} + O(\epsilon) .
\end{equation}

\begin{figure}
\begin{center}
\includegraphics[width=80mm]{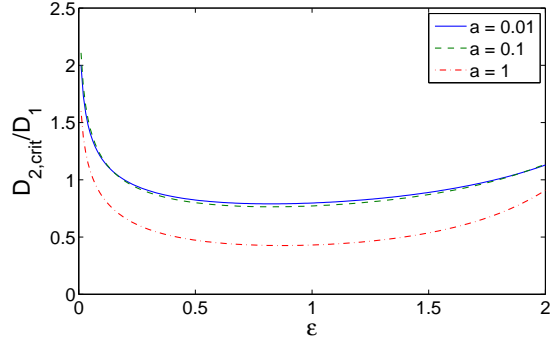}
\end{center}
\caption{
$D_{2,\rm crit}$ as a function of $\epsilon$ is computed from
Eq. (\ref{eq:D2crit_3d}) in 3D for three values of $a/R$: $0.01$,
$0.1$, and $1$.  When $\epsilon$ approaches $\pi$ (the whole surface
becomes absorbing), $D_{2,\rm crit}$ diverges (not shown).  In fact,
in this limit, there is no need for a bulk excursion because the
target will be found immediately by the surface diffusion.  In
addition, $D_{2,\rm crit}$ also diverges as $\epsilon\to 0$ because a
point-like target cannot be detected neither by bulk excursions, nor
by surface diffusion in 3D. }
\label{fig:D2crit_3d}
\end{figure}

\begin{figure}
\begin{center}
\includegraphics[width=80mm]{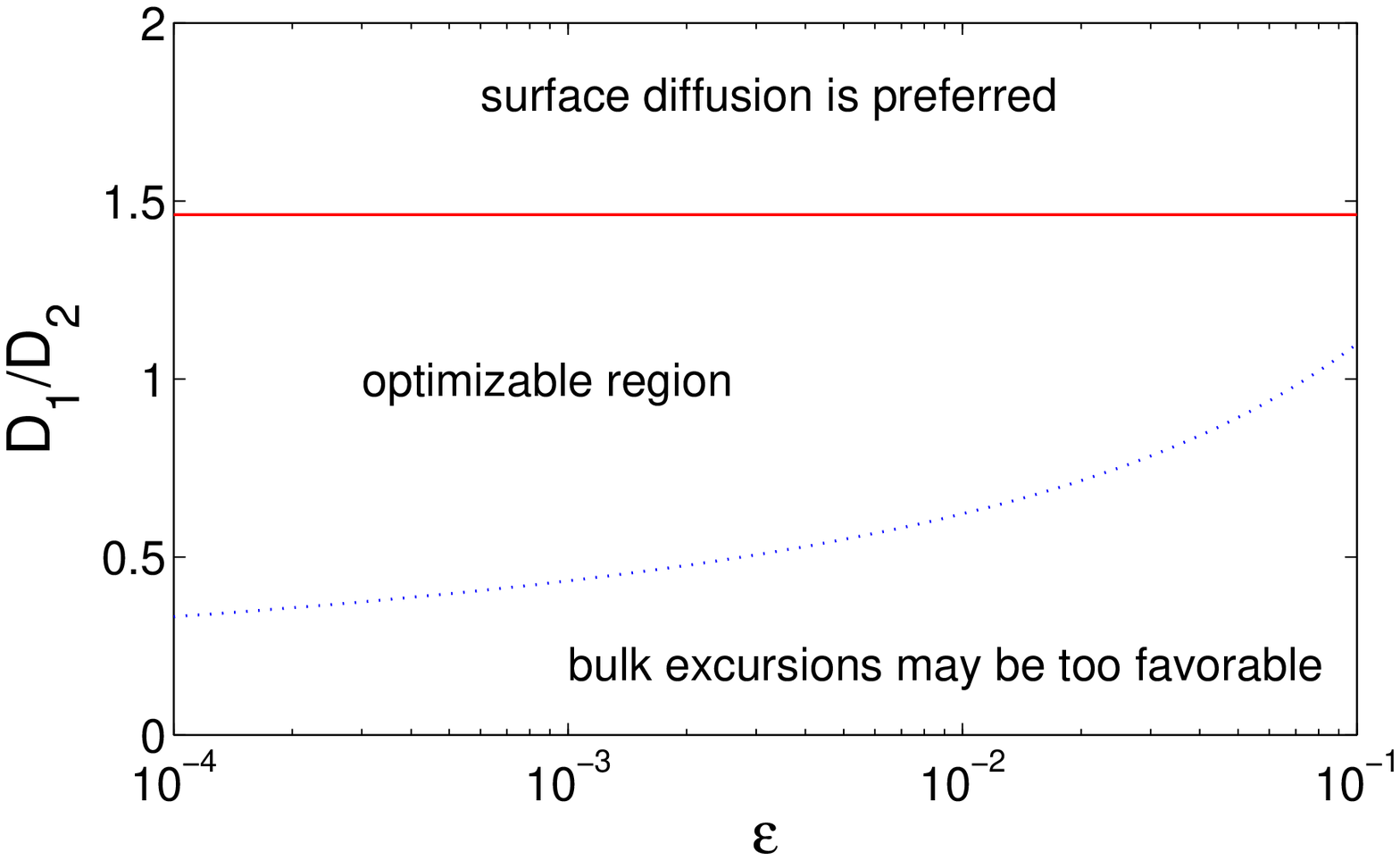}
\includegraphics[width=80mm]{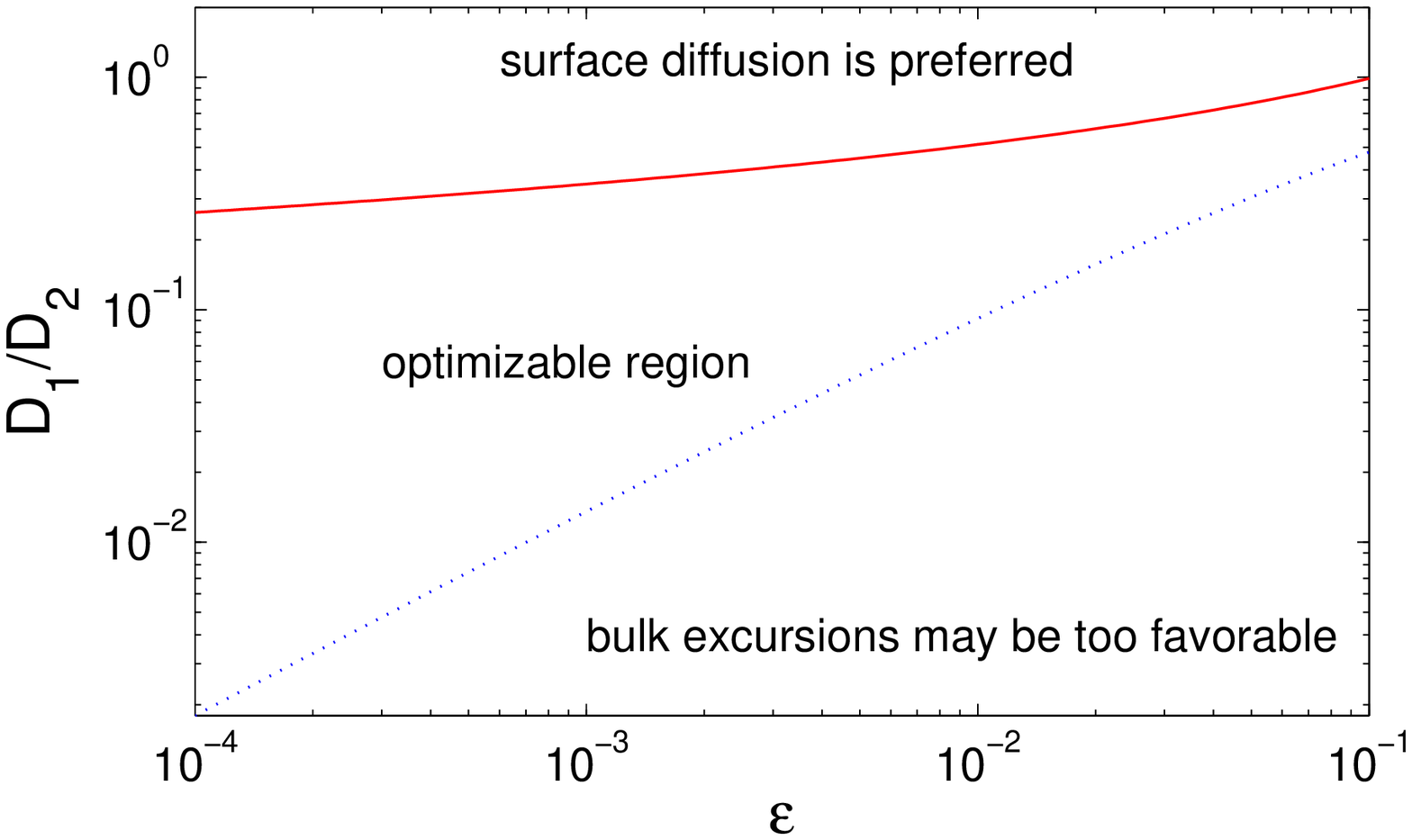}
\end{center}
\caption{
The regions of optimality for the search time in 2D (left) and 3D
(right).  The lower bound (solid blue line) and upper bound (dashed
red line) are given in the limit of $a\ll R$ and $\epsilon \ll 1$ by
Eqs. (\ref{eq:optim_region_2d}, \ref{eq:optim_region_3d}) in 2D and
3D, respectively.  When the ratio $D_1/D_2$ lies between two curves,
the search time $\langle t_1\rangle$ is optimizable with respect to
$\lambda$.  Above the upper bound, surface diffusion is preferred
($\lambda = 0$ is the optimal solution), while below the lower bound,
bulk excursions may be "too favorable" ($\lambda\to \infty$ may give the
optimal solution).  We recall that the lower bound was obtained from
the sufficient condition (\ref{eq:cond_suff}) meaning that the region
below the dotted line may still be optimizable. }
\label{fig:optimizable}
\end{figure}

There are similarities and differences between the behaviors of
$D_{2,\rm crit}$ in 2D and 3D.  Figure \ref{fig:D2crit_3d} shows that
$D_{2,\rm crit}$ from Eq. (\ref{eq:D2crit2_3d}) is not a monotonous
function of $\epsilon$, with the qualitative explanation which is the
same as in the two-dimensional case.

In contrast to the analogous Eq. (\ref{eq:D2crit2_2d}) in 2D, the rhs
of Eq. (\ref{eq:D2crit2_3d}) diverges as $\epsilon\to 0$.  This
divergence reflects the fact that a poink-like target ($\epsilon =
0$), which could be found within a finite time in 2D by
one-dimensional surface diffusion on the circle, is not detectable in
3D neither by bulk excursions, nor by surface diffusion.

\subsubsection{When is there an optimal value of the desorption rate $\lambda$ minimizing the search time ?}

For the reaction time to be an optimizable function of the desorption
rate $\lambda$, it is necessary to write an additional condition,
requiring that the bulk excursions are not "too favorable" (otherwise,
the best strategy is obtained for $\lambda\to\infty$).  A sufficient
condition is given by demanding that the search time at zero
desorption rate (i.e., without leaving the boundary) is less than the
search time at infinite desorption rate
\begin{equation}
\label{eq:cond_suff}
\langle t_1(\lambda=0) \rangle < \langle t_1(\lambda\to\infty)\rangle .
\end{equation}

\begin{equation}
\langle t_1 (\lambda=0) \rangle = \frac{R^2}{D_1}(2\ln(2/\epsilon)-1) + O(\epsilon),
\end{equation}
which writes in the physically relevant limit $a\ll R$ (using the
result of \cite{Singer:2006a}):
\begin{equation}
\langle t_1 (\lambda\to\infty) \rangle= \frac{\pi R^2}{3\epsilon D_2}\left(1+\epsilon\ln(1/\epsilon)\right).
\end{equation}
Finally, this conditions leads, for small $\epsilon$,  to
\begin{equation}
\label{cond4}
\frac{D_1}{D_2}>\frac{3\epsilon}{\pi}(2\ln(2/\epsilon)-1).
\end{equation}
Combining the two conditions (\ref{eq:D2crit2_3d}, \ref{cond4}), the
search time is found to be optimizable when $a\ll R$ and $\epsilon\ll
1$ if
\begin{equation}
\label{eq:optim_region_3d}
\frac{3\epsilon}{\pi}(2\ln(2/\epsilon)-1)<\frac{D_1}{D_2}<\frac{\pi^2}{2(2\ln(2/\epsilon)-1)} .
\end{equation}

\section{Numerical resolution}
\label{numerics}

In the previous sections, we derived the closed matrix forms
(\ref{eq:dn_2d}, \ref{eq:dn_3d}) for the coefficients $d_n$ in
2D and  3D.  These coefficients determine the angular
dependence of $t_1(\theta)$ through the explicit representations
(\ref{eq:psi_2d}, \ref{eq:psi_3d}) in 2D and 3D, respectively.
Although the formulas (\ref{eq:dn_2d}, \ref{eq:dn_3d}) which are
based on the inversion of an infinite-dimensional matrix $(I-\Omega
Q)$ remain implicit, a numerical resolution of the problem has become
straightforward.  In fact, one needs to truncate the
infinite-dimensional matrix $Q$ and vectors $U$ and $\tilde{U}$ and to
invert the truncated matrix $(I-\Omega Q)$ numerically.  

There are six parameters that determine the function $t_1(\theta)$:
the radius $R$ of the disk (sphere), the diffusion coefficients $D_1$
and $D_2$, the desorption rate $\lambda$, the size $\epsilon$ of the
absorbing region, and the distance $a$.  From now on, we set the units
of length and time by setting $R = 1$ and $D_1 = 1$.  Although the
distance $a$ may take any value from $0$ to $R$, the physically
interesting case corresponds to the limit of small $a$.  As we
mentioned previously, the limit $a = 0$ exists but trivially leads to
searching on the surface, without intermediate bulk excursions.  In
order to reveal the role of $a$, we consider several values of $a$:
$0.001$, $0.01$, $0.1$ and $1$, the latter corresponding to the
specific situation when search is always restarted from the center.
Since the diffusion coefficient $D_2$ enters only through the
prefactor $T$ from Eq. (\ref{defT}), its influence onto the searching
time $t_1$ is easy to examine.  In what follows, we take three values
of $D_2$: $0.1$, $1$ and $10$.  The dependence of $t_1$ on the
desorption rate $\lambda$ and the size $\epsilon$ is the most
interesting issue which will be studied below.

In the previous sections, we derived several formulas for computing
$t_1$: 
\begin{itemize}
\item
explicit representations (\ref{eq:psi_2d}, \ref{eq:psi_3d}) with the
exact expressions (\ref{eq:dn_2d}, \ref{eq:dn_3d}) for the
coefficients;

\item
approximations (\ref{eq:psi_part_2d}, \ref{eq:psi_3d_part}) which were
derived by neglecting non-diagonal elements of the matrix $Q$;

\item
perturbative formulas (\ref{eq:alphanperturbation0},
\ref{eq:psi_3d_perturb}) which are valid for small $\epsilon$.

\end{itemize}

For a numerical computation of the coefficients in
Eqs. (\ref{eq:dn_2d}, \ref{eq:dn_3d}), we truncate the
infinite-dimensional matrix $Q$ to a finite size $N\times N$ and
invert the matrix $(I-\Omega Q)$.  In order to check the accuracy of
this scheme, we compute the coefficients by taking several values of
$N$ from $10$ to $200$.  For $D_2 = 1$, $\epsilon = 0.1$, $a = 0.01$
and $\lambda = 1$, the computed mean time $\langle t_1 \rangle$
rapidly converges to a limit.
Even the computation with $N = 10$ gives the result with four
significant digits.  Note that other sets of parameters (e.g., larger
values of $\Omega$) may require larger truncation sizes.


\section{Conclusion}

To conclude, we have presented an exact calculation of the mean
first-passage time to a target on the surface of a 2D and 3D spherical
domain, for a molecule performing surface-mediated diffusion.  The
presented approach is based on an integral equation which can be
solved analytically, and numerically validated approximation schemes,
which provide more tractable expressions of the mean FPT.  This
minimal model of surface-mediated reactions, which explicitly takes
into account the combination of surface and bulk diffusions, shows the
importance of correlations induced by the coupling of the switching
dynamics to the geometry of the confinement.  Indeed, standard MF
treatments prove to substantially underestimate the reaction time in
this case \cite{Obenichou:2008}, and sometimes even fail to reproduce
the proper monotonicity \cite{Oshanin:2010}.  In the context of interfacial
systems in confinement, our results show that the reaction time can be
minimized as a function of the desorption rate from the surface, which
puts forward a general mechanism of enhancement and regulation of
chemical reactivity.

\appendix
\section{Another approach in 2D }
\label{sec:approach}

In this Appendix, we describe another theoretical approach which
relies on the explicit form of the Green function of the Poisson
equation in 2D case.  In particular, the perturbative analysis for
small $\epsilon$ becomes easier within this approach.

Considering $t_2$ as a source term in the Poisson type equation
(\ref{eq:t1}) with absorbing conditions at $\theta=\epsilon$ and
$\theta=2\pi-\epsilon$ whose Green function is well known
\cite{Barton:1989}, $t_1$ writes
\begin{eqnarray}
\label{eq:t1integre}
t_1(\theta)&=&\frac{1}{\omega \sinh(2 \omega(\pi-\epsilon))}\int _\epsilon ^{2\pi-\epsilon} 
\sinh(\omega (\theta_<-\epsilon))\sinh (\omega(2\pi - \epsilon-\theta_>))
\left[\frac{R^2}{D_1}+\frac{\lambda R^2}{D_1} t_2(R-a,\theta')\right]{\rm d} \theta',
\end{eqnarray}
and the notations $\theta_<=\min(\theta,\theta')$ and
$\theta_>=\max(\theta,\theta')$.

Injecting Eq. (\ref{eq:t2integre}) into Eq. (\ref{eq:t1integre}) leads to
\begin{equation}
t_1(\theta)=\frac{\omega}{\lambda \sinh (2\omega(\pi-\epsilon))}\left({\rm I}(0,\theta)
\left(1+\lambda\left(\alpha_0-\frac{(R-a)^2}{4D_2}\right)\right)+\lambda\sum_{m=1}^\infty \alpha_m(R-a)^m{\rm I}(m,\theta)\right),
\end{equation}
where, for $m$ integer, 
\begin{eqnarray}
{\rm I}(m,\theta)&\equiv& \int_\epsilon^{2\pi -\epsilon}\sinh(\omega(\theta_<-\epsilon))
\sinh(\omega(2\pi-\epsilon-\theta_>))\cos(m\theta'){\rm d}\theta'\nonumber\\
&=&\frac{\omega}{\omega^2+m^2}\left(\cos(m\theta)\sinh(2\omega(\pi-\epsilon))-2\cos(m\epsilon)
\sinh(\omega(\pi-\epsilon))\cosh(\omega(\theta-\pi))\right),
\end{eqnarray}
so that 
\begin{eqnarray}
\label{eq:t1complet}
t_1(\theta)&=&\frac{1}{\lambda}+\alpha_0-\frac{(R-a)^2}{4D_2}+
\omega^2\sum_{m=1}^\infty \frac{\alpha_m}{\omega^2+m^2}(R-a)^m\cos(m\theta)
\nonumber\\
&-&\frac{\cosh(\omega(\theta-\pi))}{\cosh(\omega(\pi-\epsilon))}\left(\frac{1}{\lambda}+
\alpha_0-\frac{(R-a)^2}{4D_2}+\omega^2\sum_{m=1}^\infty \frac{\alpha_m}{\omega^2+m^2}(R-a)^m\cos(m\epsilon)\right).
\end{eqnarray}

Substituting Eq. (\ref{eq:t1complet}) into Eq. (\ref{eq:defFourier}) gives
\begin{eqnarray}
\label{eq:mode0}
S\frac{\tanh(\omega(\pi-\epsilon))}{\omega\pi}=-\frac{\epsilon}{\pi}\lambda \alpha_0 +1 -
\frac{\epsilon}{\pi}+\lambda\left(\frac{R^2}{4D_2}-\left(1-\frac{\epsilon}{\pi}\right)\frac{(R-a)^2}{4D_2}\right)
-\frac{\lambda\omega^2}{\pi}\sum_{m=1}^\infty\frac{\alpha_m}{\omega^2+m^2}(R-a)^m\frac{\sin(m\epsilon)}{m},
\end{eqnarray}
and
\begin{eqnarray}
\label{eq:moden}
&&\lambda\left(R^n-\frac{\omega^2}{\omega^2+n^2}\left(1-\frac{\epsilon}{\pi}\right)(R-a)^n\right)\alpha_n=\nonumber\\
&-&\frac{2\sin(n\epsilon)}{n\pi}\left(1+\lambda\left(\alpha_0-\frac{(R-a)^2}{4D_2}\right)\right)\nonumber\\
&-&\frac{\lambda\omega^2}{\pi}\left(\sum_{m\neq n}\frac{\alpha_m}{\omega^2+m^2}(R-a)^m\frac{\sin((m-n)\epsilon)}{m-n} 
+ \sum_{m=1}^\infty\frac{\alpha_m}{\omega^2+m^2}(R-a)^m\frac{\sin((m+n)\epsilon)}{m+n}\right)\nonumber\\
&-&\frac{2\tanh(\omega(\pi-\epsilon))}{\pi(\omega^2+n^2)}S\left(\omega\cos(n\epsilon)-\frac{n\sin(n\epsilon)}{\tanh(\omega(\pi-\epsilon))}\right),
\end{eqnarray}
where
\begin{equation}
S\equiv 1+\lambda\left(\alpha_0-\frac{(R-a)^2}{4D_2}\right)+\lambda\omega^2\sum_{m=1}^\infty \frac{\alpha_m}{\omega^2+m^2}(R-a)^m \cos(m\epsilon).
\end{equation}
Eq. (\ref{eq:mode0}) can be rearranged into
\begin{eqnarray}
\label{eq:rearrangement}
(\alpha_0-\frac{R^2}{4D_2})\left(\frac{\epsilon}{\pi}+\frac{\tanh(\omega(\pi-\epsilon))}{\pi\omega}\right)&=
&\left(\frac{1}{\lambda}+\left(\frac{R^2}{4D_2}-\frac{(R-a)^2}{4D_2}\right)\right)
\left(1-\frac{\epsilon}{\pi}-\frac{\tanh(\omega(\pi-\epsilon))}{\pi\omega}\right)\nonumber\\
&-&\frac{\omega}{\pi}\sum_{n=1}^\infty \frac{(R-a)^n}{\omega^2+n^2}
\left(\tanh(\omega(\pi-\epsilon))\cos(n\epsilon)+\frac{\omega}{n}\sin(n\epsilon)\right)\alpha_n,
\end{eqnarray}
and Eq. (\ref{eq:moden}) into
\begin{eqnarray}
\label{eq:moden2}
&&\left(R^n-\frac{\omega^2}{\omega^2+n^2}\left(1-\frac{\epsilon}{\pi}\right)(R-a)^n\right)\alpha_n=\nonumber\\
&&-\frac{2}{\pi n}(\alpha_0-\frac{R^2}{4D_2}+T)\left(\frac{\omega^2}{\omega^2+n^2}\sin(n\epsilon) + 
\frac{n\omega}{\omega^2+n^2}\tanh(\omega(\pi-\epsilon))\cos(n\epsilon)\right)\nonumber\\
&&-\frac{2\omega^2}{\pi (\omega^2+n^2)}\sum_{m=1}^\infty \alpha_m \frac{(R-a)^m}{\omega^2+m^2}\cos(m\epsilon)
(\omega\cos(n\epsilon)\tanh(\omega(\pi-\epsilon))-n\sin(n\epsilon))\nonumber\\
&&-\frac{\omega^2}{\pi}\left(\sum_{m\neq n}\frac{\alpha_m}{\omega^2+m^2}(R-a)^m\frac{\sin((m-n)\epsilon)}{m-n} + 
\sum_{m=1}^\infty\frac{\alpha_m}{\omega^2+m^2}(R-a)^m\frac{\sin((m+n)\epsilon)}{m+n}\right).
\end{eqnarray}

\subsection{Particular case $\lambda=0$}
 
In this case, the previous equations can be solved exactly, leading to 
\begin{equation}
\alpha_0-\frac{R^2}{4D_2}=\frac{1}{3}\frac{R^2}{D_1}\frac{(\pi-\epsilon)^3}{\pi},
\end{equation}
and
\begin{equation}
\alpha_n=-\frac{2}{\pi}\frac{R^2}{D_1}\frac{n(\pi-\epsilon)\cos(n\epsilon)+\sin(n\epsilon)}{n^3}\frac{1}{R^n} .
\end{equation}
We note that the particular case $a=0$ is also described by these
expressions, although it does not seem to be clear from
Eqs. (\ref{eq:mode0})-(\ref{eq:moden}).

\subsection{Particular case $a=R$}
 
Here again, Eqs. (\ref{eq:mode0})-(\ref{eq:moden}) can be solved exactly,
and give : 
\begin{equation}
\alpha_0-\frac{R^2}{4D_2}=\left(\frac{1}{\lambda}+\frac{R^2}{4D_2}\right)\frac{1-\frac{\epsilon}{\pi}-
\frac{\tanh(\omega(\pi-\epsilon))}{\pi\omega}}{\frac{\epsilon}{\pi}+\frac{\tanh(\omega(\pi-\epsilon))}{\pi\omega}},
\end{equation}
and
\begin{equation}
\alpha_n=-\frac{2}{\pi}\left(\frac{1}{\lambda}+\frac{R^2}{4D_2}\right)\frac{\omega}{\omega^2+n^2}
\frac{\frac{\omega}{n}\sin(n\epsilon)+\tanh(\omega(\pi-\epsilon))\cos(n\epsilon)}{\frac{\epsilon}{\pi}+
\frac{\tanh(\omega(\pi-\epsilon))}{\pi\omega}}\frac{1}{R^n} .
\end{equation}

\subsection{Perturbative approach}

Expanding $\alpha_0$ and $\alpha_n$ in powers of $\epsilon$
\begin{equation}
\alpha_0=\alpha_0^{(0)}+\alpha_0^{(1)}\epsilon+\alpha_0^{(2)}\epsilon^2+\dots\;\;{\rm and}\;\; 
\alpha_n=\alpha_n^{(0)}+\alpha_n^{(1)}\epsilon+\alpha_n^{(2)}\epsilon^2+\dots
\end{equation}
Eqs. (\ref{eq:mode0})-(\ref{eq:moden}) lead, after lengthy calculations, to
\begin{equation}
\label{eq:alphanperturbation}
\begin{split}
\alpha_0 & = \frac{R^2}{4D_2} + \omega^2T \left\{ \left(2\sum_{m=1}^\infty\frac{1}{\omega^2\left(1-x^m\right)+m^2}\right)
-\pi \epsilon+ \left(1+2\omega^2\sum_{m=1}^\infty \frac{1-x^m}{\omega^2\left(1-x^m\right)+m^2}\right)\epsilon^2\right\}+\dots ,\\
\alpha_n & = \frac{\omega^2 T}{R^n(\omega^2 (1-x^n) + n^2)}\left\{
-2+n^2\epsilon^2+\dots \right\} .
\end{split}
\end{equation}

\section{A second integral equation satified by $t_1$ in the 2D case}

Using Eq. (\ref{eq:defFourier}) for the Fourier coefficients in
Eq. (\ref{eq:t1complet}) leads to a second integral equation satisfied
by $t_1$
\begin{eqnarray}
t_1(\theta)=T\left(1-\frac{\cosh(\omega(\pi-\theta))}{\cosh(\omega(\pi-\epsilon))}\right) + 
\int_\epsilon^{2\pi-\epsilon}t_1(\alpha)\left(J(\theta,\alpha)-\frac{\cosh(\omega(\pi-\theta))}{\cosh(\omega(\pi-\epsilon))}
J(\epsilon,\alpha)\right){\rm d}\alpha,
\end{eqnarray}
where
\begin{equation}
J(\theta,\alpha)\equiv \frac{1}{2\pi}+\frac{1}{\pi}\sum_{n=1}^\infty \frac{\omega^2}{\omega^2+n^2}
\left(1-\frac{a}{R}\right)^n\cos(n\theta) \cos(n\alpha).
\end{equation}
This equation is especially well adapted to local expansions of
$t_1(\theta)$ in the vicinity of $a\simeq R$, but it can also be
rearranged into the following integral equation, useful when $a\ll R$
:
\begin{eqnarray}
t_1(\theta)&=&T\left(1-\frac{\cosh(\omega(\pi-\theta))}{\cosh(\omega(\pi-\epsilon))}\right) + 
\frac{\omega}{\sinh(2\omega (\pi-\epsilon))} \int_\epsilon^{2\pi-\epsilon}t_1(\theta')\sinh(\omega (\theta_<-\epsilon))
\sinh (\omega(2\pi - \epsilon-\theta_>)){\rm d}\theta'+\nonumber\\
&+&\int_\epsilon^{2\pi-\epsilon}t_1(\alpha)\left(\widetilde{J}(\theta,\alpha)-\frac{\cosh(\omega(\pi-\theta))}
{\cosh(\omega(\pi-\epsilon))}\widetilde{J}(\epsilon,\alpha)\right){\rm d}\alpha,
\end{eqnarray}
where
\begin{equation}
\widetilde{J}(\theta,\alpha)\equiv \frac{1}{\pi}\sum_{n=1}^\infty \frac{\omega^2}{\omega^2+n^2}
\left(\left(1-\frac{a}{R}\right)^n-1\right) \cos(n\theta) \cos(n\alpha).
\end{equation}

\section{Computation of $I_\epsilon(m,n)$ in 3D }
\label{sec:A_I}

In this Appendix, we provide the explicit formula for the matrix
$I_\epsilon(m,n)$ in 3D case.  Although technical, this is an
important result for a numerical computation because it allows one to
avoid an approximate integration in Eq. (\ref{eq:I_3d}) which
otherwise could be a significant source of numerical errors.  The
formula (\ref{eq:PmPn}) for non-diagonal elements is somewhat
elementary, while the derivation for diagonal elements seems to be
original.

\subsection*{ Non-diagonal elements }

The Legendre polynomials satisfy
\begin{equation}
\frac{d}{dx} \biggl[(1-x^2) \frac{d}{dx} P_n(x)\biggr] + n(n+1)P_n(x) = 0 ,
\end{equation}
from which 
\begin{equation}
\int\limits_a^b dx P_n(x) = - \frac{\bigl[(1-x^2) P'_n(x)\bigr]_a^b}{n(n+1)}  \hskip 5mm (n > 0) .
\end{equation}
and
\begin{equation}
\int\limits_a^b dx P_m(x) P_n(x) = \frac{\bigl[(1-x^2) [P_m(x) P'_n(x) - P_n(x) P'_m(x)]\bigr]_a^b}{m(m+1) - n(n+1)}
\hskip 5mm (m \ne n) .
\end{equation}

Since
\begin{equation}
(1-x^2) P'_n(x) = -n x P_n(x) + nP_{n-1}(x) = (n+1) x P_n(x) - (n+1) P_{n+1}(x) ,
\end{equation}
we find
\begin{equation}
\int\limits_a^b dx P_n(x) = \frac{\bigl[x P_n(x) - P_{n-1}(x) \bigr]_a^b}{n+1}  \hskip 5mm (n > 0) 
\end{equation}
and
\begin{equation}
\label{eq:PmPn}
\int\limits_a^b dx P_m(x) P_n(x) = \frac{\bigl[(m-n)x P_m(x) P_n(x) + n P_{n-1}(x) P_m(x) - m P_{m-1}(x) P_n(x)]\bigr]_a^b}{m(m+1) - n(n+1)}
\hskip 5mm (m \ne n).
\end{equation}

From the above formulas, we get
\begin{equation}
\begin{split}
I_\epsilon(m,n) & = m \frac{(n-m) u P_m(u) P_n(u) + (m+1)P_m(u) P_{n-1}(u) - (n+1)P_n(u)P_{m-1}(u)}{(n+1)[m(m+1)-n(n+1)]} , \\
 u & = \cos \epsilon  ~~~ (m \ne n). \\
\end{split}
\end{equation}

\subsection*{ Diagonal elements }

We denote
\begin{equation}
K_n = \int\limits_a^b dx P_n^2(x) .
\end{equation}
Using the relation
\begin{equation}
P_n(x) = \frac{2n-1}{n} x P_{n-1}(x) - \frac{n-1}{n} P_{n-2}(x) ,
\end{equation}
we obtain
\begin{equation}
K_n = \frac{2n-1}{n} \int\limits_a^b dx x P_{n-1}(x) P_n(x) - \frac{n-1}{n} \int\limits_a^b dx P_{n-2}(x) P_n(x) .
\end{equation}
The second integral is given by Eq. (\ref{eq:PmPn}).  In order to
compute the first one, we consider
\begin{equation}
\begin{split}
0 & = \int\limits_a^b dx \biggl\{x P_{n-1}(x) \biggl[\frac{d}{dx} \biggl[(1-x^2) \frac{d}{dx} P_n(x)\biggr] + n(n+1)P_n(x)\biggr] \\
  & - x P_n(x) \biggl[\frac{d}{dx} \biggl[(1-x^2) \frac{d}{dx} P_{n-1}(x)\biggr] + (n-1)nP_{n-1}(x)\biggr]\biggr\} \\
  & = 2n \int\limits_a^b dx x P_{n-1}(x) P_n(x) + \biggl[xP_{n-1}(x) (1-x^2) P'_n(x) - xP_n(x) (1-x^2) P'_{n-1}(x)\biggr]_a^b  \\
  & - \int\limits_a^b dx (1-x^2) \bigl[ P'_n(x) P_{n-1}(x) - P'_{n-1}(x) P_n(x)\bigr] . \\
\end{split}
\end{equation}
The last integral can be written as
\begin{equation}
\begin{split}
J & = \int\limits_a^b dx (1-x^2) P_n^2(x) (P_{n-1}(x)/P_n(x))' = \bigl[(1-x^2) P_{n-1}(x) P_n(x) \bigr]_a^b \\
& - \int\limits_a^b dx (P_{n-1}(x)/P_n(x)) \bigl[-2x P_n^2(x) + 2(1-x^2) P'_n(x) P_n(x)\bigr] \\
& = \bigl[(1-x^2) P_{n-1}(x) P_n(x) \bigr]_a^b - 
2\int\limits_a^b dx \bigl[-x P_{n-1}(x)P_n(x) + (1-x^2) P'_n(x) P_{n-1}(x)\bigr] . \\
\end{split}
\end{equation}
In the last term, we substitute $(1-x^2)P'_n(x)$ to get
\begin{equation}
J = \bigl[(1-x^2) P_{n-1}(x) P_n(x) \bigr]_a^b + 2\int\limits_a^b dx x P_{n-1}(x)P_n(x) 
- 2\int\limits_a^b dx \bigl[-nxP_n(x) + nP_{n-1}(x)\bigr] P_{n-1}(x) .
\end{equation}
Bringing these results together, we get
\begin{equation}
\begin{split}
0 & = 2n \int\limits_a^b dx x P_{n-1}(x) P_n(x) + \biggl[xP_{n-1}(x) (1-x^2) P'_n(x) - xP_n(x) (1-x^2) P'_{n-1}(x)\biggr]_a^b  \\
  & + \bigl[(1-x^2) P_{n-1}(x) P_n(x) \bigr]_a^b + 2\int\limits_a^b dx x P_{n-1}(x)P_n(x) - 
2\int\limits_a^b dx \bigl[-nxP_n(x) + nP_{n-1}(x)\bigr] P_{n-1}(x) \\
\end{split}
\end{equation}
so that
\begin{equation}
\int\limits_a^b dx x P_{n-1}(x) P_n(x)  = \frac{-1}{4n+2} \biggl[x (1-x^2) \bigl[P_{n-1}(x)P'_n(x) - P_n(x) P'_{n-1}(x)] 
+ (1-x^2) P_{n-1}(x) P_n(x)\biggr]_a^b + \frac{n}{2n+1} K_{n-1} .
\end{equation}

We obtain
\begin{equation}
\begin{split}
K_n & = - \frac{2n-1}{2n(2n+1)} \biggl[x (1-x^2) \bigl[P_{n-1}(x)P'_n(x) - P_n(x) P'_{n-1}(x)] + 
(1-x^2) P_{n-1}(x) P_n(x)\biggr]_a^b + \frac{2n-1}{2n+1} K_{n-1} \\
    & - \frac{n-1}{n} \biggl[\frac{2xP_{n-2}(x)P_n(x) - nP_{n-1}(x)P_{n-2}(x) + (n-2)P_{n-3}(x)P_n(x)}{2(2n-1)}\biggr]_a^b . \\
\end{split}
\end{equation}
We can further simplify this expression by using the following
identities
\begin{equation}
\begin{split}
& (n-1)P_{n-1}(x) - (2n-3)x P_{n-2}(x) + (n-2) P_{n-3}(x) = 0, \\
& nP_n(x) - (2n-1)x P_{n-1}(x) + (n-1) P_{n-2}(x) = 0, \\
& (1-x^2) P'_n(x) = -nxP_n(x) + nP_{n-1}(x), \\
& (1-x^2) P'_{n-1}(x) = nx P_{n-1}(x) - nP_n(x) . \\
\end{split}
\end{equation}
We get
\begin{equation}
\begin{split}
K_n & = - \frac{2n-1}{2n(2n+1)} \biggl[nx \bigl[P_{n-1}^2(x) + P_n^2(x) - 2x P_n(x) P_{n-1}(x)] + 
(1-x^2) P_{n-1}(x) P_n(x)\biggr]_a^b + \frac{2n-1}{2n+1} K_{n-1} \\
    & - \frac{n-1}{2n(2n-1)} \biggl[(2n-1)x P_n(x)P_{n-2}(x) - nP_{n-1}(x)P_{n-2}(x) - (n-1)P_{n-1}(x)P_n(x)\biggr]_a^b \\
    & = - \frac{2n-1}{2n(2n+1)} \biggl[nx \bigl[P_{n-1}^2(x) + P_n^2(x) - 2x P_n(x) P_{n-1}(x)] + 
(1-x^2) P_{n-1}(x) P_n(x)\biggr]_a^b + \frac{2n-1}{2n+1} K_{n-1} \\
    & - \frac{1}{2n} \biggl[((2n-1)x^2 + 1) P_n(x)P_{n-1}(x) - nx (P_{n-1}^2(x) + P_n^2(x))\biggr]_a^b \\
    & = \frac{\bigl[x (P_{n-1}^2(x) + P_n^2(x)) - 2P_n(x) P_{n-1}(x)\bigr]_a^b}{2n+1}  + \frac{2n-1}{2n+1} K_{n-1} \\
\end{split}
\end{equation}
and we know that $K_0 = b-a$.  Applying this formula recursively, one
finds
\begin{equation}
K_n = \frac{F_n(b) - F_n(a)}{2n+1} ,
\end{equation}
where
\begin{equation}
\begin{split}
F_n(x) & = x[P_n^2(x) + 2P_{n-1}^2(x) + ... + 2P_1^2(x) + P_0(x)] \\
& - 2P_n(x)P_{n-1}(x) - 2P_{n-1}(x)P_{n-2}(x) - ... - 2P_1(x)P_0(x) + x \\
       & = \sum\limits_{k=1}^n \bigl[2(x-1)P_k^2(x) + [P_k(x) - P_{k-1}(x)]^2\bigr] - (x-1)P_n^2(x) + (x-1)P_0^2(x) + x .\\
\end{split}
\end{equation}
One can check that this function satisfies the recurrent relation
\begin{equation}
F_n(x) = F_{n-1}(x) + x[P_n^2(x) + P_{n-1}^2(x)] - 2P_n(x) P_{n-1}(x) , \hskip 5mm  F_0(x) = x . 
\end{equation}
Note that $F_n(\pm 1) = F_{n-1}(\pm 1) = ... = \pm 1$.

As a result, we obtain
\begin{equation}
I_\epsilon(n,n) = - P_n(u) \frac{u P_n(u) - P_{n-1}(u)}{n+1} + \frac{F_n(u) + 1}{2n+1},  \hskip 5mm  u = \cos\epsilon .
\end{equation}

\bibliographystyle{vincent}


\end{document}